\documentclass{aa}  
\usepackage[utf8]{inputenc}

\usepackage[switch]{lineno}

\usepackage{graphicx}

\usepackage{txfonts}
\usepackage{hyperref}
\usepackage{ltablex}
\usepackage{changepage}
\usepackage{booktabs} 
\usepackage{caption}
\usepackage{appendix} 
\usepackage{xcolor}
\usepackage{float}
\usepackage{adjustbox}

\hypersetup{
    colorlinks=true,
    linkcolor=blue,  
    citecolor=blue, 
    urlcolor=blue     
}

\usepackage{booktabs}
\usepackage{verbatim} 
\usepackage{amsmath}

\usepackage{threeparttable}
\usepackage{longtable}

\usepackage{booktabs}

\usepackage{subfigure}
\makeatletter

\renewcommand{\thefigure}{\arabic{figure}}  
\renewcommand{\fnum@figure}{\textbf{Fig.~\thefigure}}  

\renewcommand{\thetable}{\arabic{table}} 
\renewcommand{\fnum@table}{\textbf{Table~\thetable}}

\renewcommand{\caption@makecaption}[2]{
    \ifx\@captype\figure
        \textbf{Fig.~\thefigure} \textbf{#2}\par  
    \else
        \textbf{Table~\thetable}: \textbf{#2}\par  
    \fi
}
\makeatother

\newcommand{\kms}{\mbox{km\,s$^{-1}$}}

\newcommand{\OIII}{\mbox{[\ion{O}{iii}]$\lambda 5007$\,\AA}}

\newcommand{\OI}{\mbox{[\ion{O}{i}]$\lambda\,6300$\,\AA}}

\newcommand{\SII}{\mbox{[\ion{S}{ii}]$\lambda\lambda\,6717, 6731$}\,\AA}

\newcommand{\lya}{\relax \ifmmode {\mbox Ly}\alpha\else Ly$\alpha$\fi}
\newcommand{\ha}{\relax \ifmmode {\mbox H}\alpha\else H$\alpha$\fi}
\newcommand{\hg}{\relax \ifmmode {\mbox H}\gamma\else H$\gamma$\fi}
\newcommand{\hd}{\relax \ifmmode {\mbox H}\delta\else H$\delta$\fi}
\newcommand{\hb}{\relax \ifmmode {\mbox H}\beta\else H$\beta$\fi}
\newcommand{\sii}{\relax \ifmmode {\mbox S\,{\scshape ii}}\else S\,{\scshape ii}\fi}
\newcommand{\nii}{\relax \ifmmode {\mbox N\,{\scshape ii}}\else N\,{\scshape ii}\fi}
\newcommand{\neiii}{\relax \ifmmode {\mbox Ne\,{\scshape iii}}\else Ne\,{\scshape iii}\fi}
\newcommand{\oii}{\relax \ifmmode {\mbox O\,{\scshape ii}}\else O\,{\scshape ii}\fi}
\newcommand{\oi}{\relax \ifmmode {\mbox O\,{\scshape i}}\else O\,{\scshape i}\fi}
\newcommand{\oiii}{\relax \ifmmode {\mbox O\,{\scshape iii}}\else O\,{\scshape iii}\fi}
\newcommand{\hii}{\relax \ifmmode {\mbox H\,{\scshape ii}}\else H\,{\scshape ii}\fi}
\newcommand{\hi}{\relax \ifmmode {\mbox H\,{\scshape ii}}\else H\,{\scshape i}\fi}

\begin{document}

   \title{Tracing ionized gas kinematics in Lyman-Break Analogs}

   \subtitle{Implications for star formation compactness and outflow properties}

   \author{León Contreras, A.
          \inst{1}\fnmsep\thanks{Corresponding authors: A. León Contreras, \email{ana.leon@userena.cl} and R. Amorín, \email{amorin@iaa.csic.es}}
    \and Amor\'in, R.\inst{2,3,1}
    \and Llerena, M. \inst{4}
    \and Fern\'andez, V.\inst{5}
          }

   \institute{Departamento de Astronom\'ia, Universidad de La Serena, Av. Juan Cisternas 1200 Norte, La Serena, Chile.\\
              \email{ana.leon@userena.cl}
    \and
        Instituto de Astrof\'{i}sica de Andaluc\'{i}a (CSIC), Apartado 3004, 18080 Granada, Spain
    \and
        Centro de Estudios de F\'{\i}sica del Cosmos de Arag\'{o}n (CEFCA), Unidad Asociada al CSIC, Plaza San Juan 1, E--44001 Teruel, Spain%2
    \and INAF - Osservatorio Astronomico di Roma, Via di Frascati 33, 00078, Monte Porzio Catone, Italy%3
    \and 
        Michigan Institute for Data Science, University of Michigan, 500 Church Street, Ann Arbor, MI 48109, USA %
    }
   \date{Received XXXXXX; accepted XXXXXX}

  \abstract
  {The ionized gas kinematics of low-mass starburst galaxies is a tracer of galaxy interactions and feedback processes, which are key for understanding massive star formation, chemical enrichment, and galaxy evolution.}
  {We study the ionized gas kinematics and outflow properties in a sample of Lyman-Break Analogs (LBAs) at $z\sim$\, 0.1-0.3, characterized by their compact morphologies, high UV luminosities, and strong emission lines, which are common at higher redshifts. }
  {We use high-resolution VLT/X-Shooter spectra of 14 compact, UV-luminous LBAs to model the complex [O{\sc iii}]$\lambda\lambda$4959,5007$\AA$ and Balmer line profiles with multi-Gaussian fits.}
  {LBAs show complex kinematics, with emission lines best reproduced by narrow ($\sigma <$\,90\,km/s) and broad ($\sigma >$90\,km/s) components in all galaxies. 
  The narrow-line kinematics is highly turbulent, likely driven by massive star-forming clumps. The luminosities and line ratios of the narrow components are typical of giant \hii\ regions. We interpret the broader components as ionized outflows driven by strong winds of massive stars and supernovae. In galaxies with highly complex profiles and disturbed morphologies, ongoing interactions or mergers are found to contribute to the broad components. We find outflow velocities in the range  $v_{\rm out}\sim$ 200 km/s to 500 km/s. Simple models yield outflow mass rates of 0.20-2.72\,$M_{\odot}\;yr^{-1}$ and mass-loading factors $\eta \sim$\,0.03-0.81.
  We find that $\eta$ shows a mild increasing trend at lower stellar masses, in agreement with previous observational studies and predictions from FIRE-2 and Illustris-TNG simulations. 
  Compact starburst morphologies may modulate the $\eta$-$M_{\star}$ relation, showing a strong $\Sigma_{SFR}$-$\eta$ correlation, i.e., more compact starbursts drive stronger outflows. 
  We find a good agreement with similar findings in star-forming galaxies at high-redshift ($z\sim$\,2-9), including those from recent JWST observations.}
  {Our results highlight the relevance of detailed studies of the ionized gas kinematics in local UV-compact starbursts to improve our understanding of feedback processes in low-mass, rapidly star-forming galaxies.}


   \maketitle

\section{Introduction}

Understanding feedback from stellar winds, supernovae (SNe), and active galactic nuclei (AGN), and its impact on the interstellar medium (ISM), is crucial to quantify the escape of ionizing photons at the epoch of reionization \citep[$z \gtrsim$\,6, ][]{Robertson_2015} and the baryon cycle that regulates galaxy growth and chemical evolution over cosmic time \citep[][]{Tumlinson2017}.
Feedback from star formation and AGN activity, together with mass and environment, is a primary driver of star-formation quenching \citep{Peng2010, Sawala2016}. 
Low-mass galaxies, with shallow potential wells, experience stronger feedback effects \citep{Muratov2015}. 
However, the impact of stellar feedback and outflows on metal enrichment and ionizing-photon escape remains uncertain in the youngest high-z galaxies \citep{llerena2023ionized,Carniani2024, deGraaff2024,RodriguezdelPino2024, Saldana-Lopez2025, Cooper2025}.

Lyman-break galaxies (LBG) are the most common population of UV-luminous star-forming galaxies at redshift $z\gtrsim$,2, largely contributing both to the galaxy number densities and the global star-formation rate density at these epochs \citep[e.g.][]{Bouwens_2021}. Thus, they are typically identified as drop-outs in deep photometric surveys by their bright and blue UV continuum with a strong depression near the Lyman limit \citep[e.g.][and references therein]{Giavalisco_2002} and later confirmed with deep spectroscopy \citep[e.g.][]{Shapley2003, LeFevre2015, Pentericci2018}. Bright LBGs typically exhibit stellar masses between \( 10^{9}-10^{10}M_{\odot} \) and star formation rates (SFRs) ranging from 10 to 100 \( M_{\odot} \) yr\(^{-1}\) \citep[]{Shapley_2001, Steidel_1996, Giavalisco_2002} and sub-solar metallicities \citep[e.g.][]{Maiolino2008, Cullen_2019}. More recent JWST observations have detected fainter LBGs with lower stellar masses at the highest redshifts \citep[]{Finkelstein2024, Arrabal_2023, Bunker2024}. However, because of their relative faintness and high redshifts, the ionized gas kinematics and outflow properties of LBGs remain challenging to study. 

Local analogs of high-$z$ LBGs offer a way to study feedback and outflows under comparable ISM conditions. 
Complementing UV-based studies with optical kinematics provides a more thorough picture of these galaxies. 
While UV absorption lines trace extended and diffuse gas on large scales, the study of ionized gas through optical emission lines provides insights into denser gas closer to regions of active star formation \citep[e.g.][]{Xu2022,Xu2025a, Martin2024}. 
A key local analog sample is the Lyman-Break Analogs \citep[LBAs][]{Heckman_2005, Hoopes_2007} at $z \sim 0.1$--0.3, defined by high FUV luminosity ($L_{\mathrm{FUV}} > 2 \times 10^{10}\,L_\odot$) and surface brightness ($I_{\mathrm{FUV}} > 10^9\,L_\odot\,\mathrm{kpc}^{-2}$).
Their multi-wavelength properties, morphologies, stellar and ISM characteristics, radio/X-ray emission, and environments have been widely studied, reinforcing their analogy with high-z galaxies.
\citep{Overzier_2008,Overzier_2009, Basu-Zych2009, Basu2009_env, Basu-Zych2013, Goncalves_2014, Contursi_2017, Loaiza, Santana_silva_2020, Santos_junior_2025, Araujo_2025}. 
Overall, they are relatively low-mass systems ($M_\star \sim 10^{9.5}$--$10^{11}\,M_\odot$) with high star formation rates (SFR $\sim 10$--$100\,M_\odot\,\mathrm{yr}^{-1}$), spanning a wide range of morphologies, with many showing evidence of interactions or merging. They also span a wide range of metallicities and ionization properties, with some of their characteristics overlapping with other local analogs, such as the Green Pea galaxies \citep[e.g.][]{Amorin_2010, Amorin2012a, Brown2014, Loaiza}.

The kinematic and feedback properties of LBAs have been studied using mostly UV absorption lines \citep{Heckman_2011}. \citet{Heckman_2015} analyzed galactic winds for a large sample of 39 LBAs using HST COS and FUSE observations and found that outflow velocity correlates weakly with stellar mass but strongly with SFR and SFR surface density. This suggests that concentrated star formation in UV-luminous galaxies drives stronger outflows. \citet{Heckman_2016} extended this analysis to extreme starburst galaxies, confirming previous trends between outflow velocity (\( V_{\text{out}} \)) and variables such as stellar mass and SFR. Their findings suggest that outflows consist of interstellar or circumgalactic clouds accelerated by a combination of gravitational forces and the momentum of the outflows. An important conclusion from the above studies is that such extreme feedback could be required to create channels or holes enabling the escape of ionizing radiation from star-forming galaxies \citet{Heckman_2011, Borthakur_2014, Alexandroff_2015}. These UV-based studies mainly trace extended, diffuse gas on galactic scales.

On the other hand, only a few studies have examined the ionized gas kinematics of LBAs through spectroscopic observations. While \citet{Overzier_2009} analyzed VLT/FLAMES high-resolution spectra of four LBAs with strong signatures of AGN activity, \citet{Goncalves_2010} presented the first spatially-resolved study of a larger sample of LBAs using AO-assisted near-infrared (NIR) integral field spectroscopy (IFS). The authors examined the most compact emission line regions of LBAs, finding relatively large velocity dispersion, suggesting strong turbulence in the ionized gas.  However, in the star-forming LBAs, they found no clear evidence of outflows, possibly due to the limited depth of such observations. Only for a few LBAs, also classified as Green Pea galaxies, \citet{Amorin_2012b} presented evidence of complex gas kinematics and ionized outflows using high-dispersion spectra. Similarly, \citet{Hogarth} studied the ionized gas kinematics of one of the most compact LBAs using echelle optical spectra, finding strong signatures of photoionized outflowing gas in the form of broad emission components in both Balmer and collisionally excited lines, similar to other local analogs \citep[][]{Bosch2019, Amorin2024}. These authors studied the physical properties and metallicity of the outflow component, finding evidence of relative enrichment due to the gas dispersed by the outflow and highlighting the role of feedback in clearing holes and channels in the ISM through which ionizing photons could escape.

Building on previous work, we analyze the ionized-gas kinematics in 14 LBAs using high-dispersion optical spectroscopy, adding new insight into their outflow properties and feedback impact. These results provide a benchmark for detailed comparisons with high-redshift LBGs. The paper is structured as follows. Section~\ref{sec2} describes the LBA sample and VLT/X-Shooter data, Section~\ref{sec3} the line-fitting methodology, Section~\ref{Result} the results, Section~\ref{sec5} the discussion, and Section~\ref{sec6} the conclusions.

Throughout this work, we assume a standard cosmological model with a Hubble constant of \( H_0 = 70 \) km s\(^{-1}\) Mpc\(^{-1}\), a matter density parameter of \( \Omega_{M} = 0.3 \), and a vacuum energy density parameter of \( \Omega_{\Lambda} = 0.7 \).

\section{Galaxy sample and data}\label{sec2}

Our sample comprises 14 LBAs, which are part of the 30 galaxies with VLT/X-Shooter spectroscopy presented in the work of \citet{Loaiza} focused on emission-line diagnostics and chemical abundances. That sample belongs to a parent sample of 74 LBAs, selected by \citet{Heckman_2005} and studied in detail by \citet{Overzier_2009}. 

The selected LBAs are relatively low-mass and compact, with stellar masses $M_{\star}\sim 10^{9.3}-10^{10.7}\; M_{\odot}$ (median $ 10^{9.9}\; M_{\odot}$) and optical effective radii 0.77–4.6\,kpc (median 1.75\,kpc) from HST F606W/F850LP imaging \citet{Overzier_2009}. Selection follows the FUV-luminosity and compactness criteria of \citet{Heckman_2005} and \citet{Hoopes_2007}. AGN or emission-line–selected systems were excluded; galaxies classified as AGN (type 1 candidates) in \citet{Loaiza} were therefore removed. One additional galaxy was excluded owing to low S/N in its X-Shooter spectra from imperfect background subtraction. Our final sample comprises 14 LBAs with high-resolution optical spectra. Table~\ref{tab:table1} presents the sample according to SDSS identifier (ID), coordinates, and spectroscopic redshift.

The spectra, originally presented by \citet{Loaiza}, were obtained with the X-Shooter spectrograph on VLT/UT2 (Kueyen).
The data are part of the program ID 085.B-0784(A) (PI: T. Heckman). 
Observations used the 11$\arcsec$ slit mode, providing simultaneous UVB (R$\sim$\,5100, 1.0$\arcsec$ slit), VIS (R$\sim$\,8900, 0.9$\arcsec$), and NIR (R$\sim$\,5100, 0.9$\arcsec$) spectra from U to K band. Total on-source integration time was 2560 s.

Compact galaxies were observed in nodding-on-slit mode for efficient sky subtraction, whereas extended ones used an offset mode with separate sky exposures. Details of the specific observing mode for each galaxy and the observing seeing can be found in \citet{Loaiza}. Seeing ranged between 1.1$\arcsec$ and 2.4$\arcsec$

The spectra were reduced by ESO, using the ESO X-Shooter reduction pipeline, EsoRex \citep{Modigliani_2010}. We downloaded calibrated (phase 3) data from the ESO archive\footnote{\url{http://archive.eso.org/scienceportal/home}}. 
Note that for the one-dimensional extraction, the pipeline considers a standard aperture of 30 pixels per side. Signal-to-noise (S/N) ratios span $\sim$10–490 for bright lines like \OIII\ and $\sim$5–70 for faint ones such as \OI. The spectral quality therefore required no further post-processing.
Additional extractions with varied apertures for extended objects confirmed the robustness of the emission-line analysis (see Sect.\ref{subsec:Extended.galaxies}).

\begin{table}[]
\centering
\caption{Information on the selected sample}
\label{tab:table1}
\begin{tabular}{llll}
\hline

ID  & \begin{tabular}[c]{@{}l@{}}R.A.\\ (J2000)\end{tabular} & \begin{tabular}[c]{@{}l@{}}Dec\\ (J2000)\end{tabular} & \begin{tabular}[c]{@{}l@{}}$z^{a}$\\ \end{tabular} \\ \hline
SDSS001009 & 00:10:09.97                                            & -00:46:03.66                                          & 0.2431  \\
SDSS004054 & 00:40:54.33                                            & 15:34:09.66                                           & 0.2832  \\
SDSS005527 & 00:55:27.46                                            & -00:21:48.71                                          & 0.1674  \\
SDSS015028 & 01:50:28.41                                            & 13:08:58.40                                           & 0.1467  \\
SDSS020356 & 02:03:56.91                                            & -08:07:58.51                                          & 0.1883  \\
SDSS021348 & 02:13:48.54                                            & 12:59:51.46                                           & 0.2190  \\
SDSS032845 & 03:28:45.99                                            & 01:11:50.85                                           & 0.1422  \\
SDSS035733 & 03:57:34.00                                            & -05:37:19.70                                          & 0.2037  \\
SDSS040208 & 04:02:08.87                                            & -05:06:42.06                                          & 0.1393  \\
SDSS143417 & 14:34:17.16                                            & 02:07:42.58                                           & 0.1803  \\
SDSS214500 & 21:45:00.26                                            & 01:11:57.58                                           & 0.2043  \\
SDSS231812 & 23:18:13.00                                            & -00:41:26.10                                          & 0.2517  \\
SDSS232539 & 23:25:39.23                                            & 00:45:07.25                                           & 0.2770  \\
SDSS235347 & 23:53:47.69                                            & 00:54:02.08                                           & 0.2234  \\ \hline
\end{tabular}
\\
\begin{flushleft}
\scriptsize{a. The redshift was obtained from the SDSS spectra by \cite{Heckman_2005}.}

\end{flushleft}
\end{table}

\section{Data analysis}\label{sec3}

\subsection{Fitting of emission lines profiles}\label{Section:Gaussian_Fits}

Inspection of the X-Shooter spectra, especially H$\alpha$  and [O{\sc iii}]$\lambda\lambda$4959,5007$\AA$, reveals distinct profiles indicating complex ionized-gas kinematics. 
We fit multiple Gaussian components to  H$\alpha$, H$\beta$, [O{\sc iii}]$\lambda\lambda$4959,5007$\AA$, [N{\sc ii}]$\lambda\lambda$6548,6584$\AA$, and [S{\sc ii}]$\lambda\lambda$6716,6731$\AA$  to characterize line shapes and derive kinematics.

This approach has been successfully employed in previous studies to delineate the kinematics of low-$z$ analogs to high-$z$ galaxies, such as  H{\sc ii} galaxies and BCDs 
 \citep[]{Chavez_2014, Melnick_1999, Firpo_2011}, Green Pea galaxies \citep[]{Amorin_2012b, Bosch2019, Hogarth}, Lyman continuum emitters \citep[LCEs, ][]{Amorin2024}, and Lyman-$\alpha$ emitters \citep[LAEs, ][]{Matthee2021,llerena2023ionized}, among others.

We use the {LiMe}\footnote{\url{https://lime-stable.readthedocs.io/en/latest/}} package \citep[]{Lime-paper}, which provides a suite of line-fitting tools for astronomical spectra. 
LiMe is built upon the \href{}{LMFIT}\footnote{\url{https://lmfit.github.io/lmfit-py/}} library \citep{lmfit}, which enables the trust-region least squares algorithm from \textsc{SciPy} \citep{2020SciPy-NMeth}. In this study, we employ nonlinear least squares fitting based on LMFIT. LiMe outputs each component’s flux amplitude, centroid, and FWHM. After fitting, LiMe provides statistical diagnostics, reduced chi-square ($\chi^2_\nu$), Akaike Information Criterion (AIC), and Bayesian Information Criterion (BIC) to assess fit quality.
Subsequently, we will briefly outline the diverse strategies and physical constraints applied to the galaxies within our sample.

\subsection{Gaussian fitting step-by-step}

We first fit [O{\sc iii}]$\lambda$5007$\r{A}$, a bright, isolated line with high S/N, ideal for initiating the fitting sequence. In our methodology, we initiate by creating a mask to define the continuum on both sides of the line and the region to fit. Subsequently, we apply a single Gaussian model with its parameters set as free variables. 
Fit quality was evaluated via visual inspection, residuals, and $\chi^2_\nu$/AIC/BIC.  Extra components were added only when a single Gaussian was inadequate. Optimal fits were selected through a combined visual and statistical assessment. 
Additional components were accepted only when $\Delta AIC = |AIC_{n-1} - AIC_{n}| > 10$ \citep{Bosch2019}. This threshold roughly corresponds to $\Delta BIC > 10$ as adopted in similar studies \citep{llerena2023ionized,Carniani2024}.

We use the kinematic parameters from [O{\sc iii}]5007 and H$\alpha$ to constrain the fits of high- and low-ionization lines, respectively. 
To do this, we use the central velocity and velocity dispersion from the bright lines to fit the faint ones. 
For doublets, theoretical flux ratios were imposed (e.g. [O{\sc iii}]4959$=$\,0.36\,$\times$\,[O{\sc iii}]5007, \citet[]{storey2000theoretical}).
H$\alpha$ and [N{\sc ii}] were fitted simultaneously, tying [\nii] centroids and widths to H$\alpha$ and fixing the theoretical ratio [\nii]6548=2.94$\times$[N{\sc ii}]6584 \citep[]{fischer2004breit}.
Similarly, [\sii]6716,6731 were fitted using H$\alpha$ kinematics and free amplitudes to derive electron densities (Sect.~\ref{sec:density}).

\subsection{The case of spatially resolved galaxies}\label{subsec:Extended.galaxies}
\begin{figure}[]
    \includegraphics[scale=0.36]{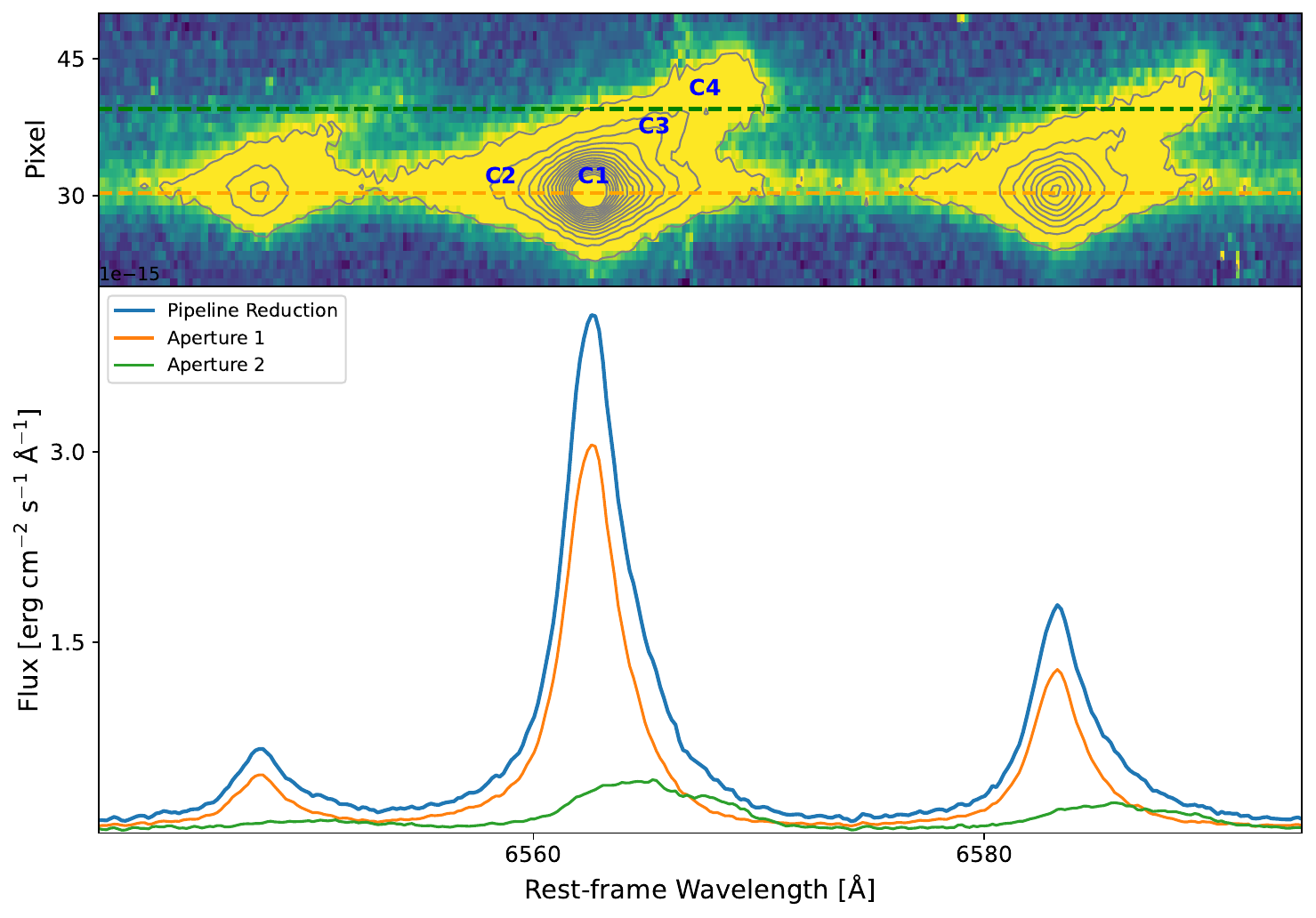}
    \caption{(Top panel) Two-dimensional spectrum of  SDSS143417 around the H$\alpha$–[\nii] region, with gray contours tracing flux levels and blue crosses marking the positions used for Gaussian fitting. (Bottom panel) One-dimensional spectra extracted with different apertures: blue (ESO pipeline, 30 px/side, including both knots), orange (IRAF/apall, 5 px centered on the brightest knot), and green (IRAF/apall, asymmetric aperture sampling the secondary knot). Vertical dashed lines indicate the wavelength ranges where the two-dimensional contours reveal asymmetries, serving as guides to identify additional kinematic components.}
    \label{all_spectrum}
\end{figure}

The above approach suits compact galaxies but requires refinement for the two spatially resolved cases, SDSS001009 and SDSS143417.
Their line profiles are complex, being spatially and spectrally resolved within the X-Shooter slit. This further complicates the determination of the central positions and the required number of components necessary for accurately fitting the emission line profiles. We therefore extracted 1-D spectra manually using IRAF tasks. For the galaxy SDSS143417, an examination of the 2D X-Shooter spectra reveals the presence of at least two spatially resolved components in possible interaction (see Figure~\ref{all_spectrum}).
The one-dimensional extraction of 30 pixels per side is optimal for most of the galaxies in the sample, except for this galaxy, for which the $H\alpha$ line profile exhibits two or more peaks. For this reason, we performed a manual one-dimensional spectrum extraction using IRAF tasks. We selected the two peaks in the spatial direction and extracted the one-dimensional spectrum accordingly. 

Figure~\ref{all_spectrum} provides a combined view of the two-dimensional and one-dimensional spectra of galaxy SDSS143417. The top panel shows contour maps of the H$\alpha$ region, revealing at least two distinct emitting regions. Because the galaxy is extended, identifying a single central position is not straightforward. To address this, we used DS9 to define contour levels and locate the centers of four Gaussian components (marked with C1, C2, C3, and C4), based on intensity maxima, minima, and asymmetries suggestive of varying gas kinematics. Colored dashed horizontal lines in the 2D spectrum mark the central positions of the spatial apertures used for the one-dimensional extractions: the orange and green dashed lines correspond to the centers of Aperture\,1 and Aperture\,2, respectively, while the central blue region represents the full pipeline extraction (labeled as "Pipeline Reduction"). This color scheme is consistent between the upper and lower panels to guide visual correspondence between the 2D and 1D spectra. The bottom panel compares one-dimensional spectra extracted using different apertures. The blue curve shows the EsoRex extraction (30 px per side), including both resolved H$\alpha$  knots in the top panel. The orange trace uses a narrow aperture on the brightest peak; the green captures the fainter knot. The comparison confirms that individual extractions reproduce the combined-spectrum features, revealing at least two spatially and spectrally resolved components.

\begin{figure*}[!ht]
    \centering
    \begin{minipage}{0.49\textwidth}
        \includegraphics[width=\linewidth]{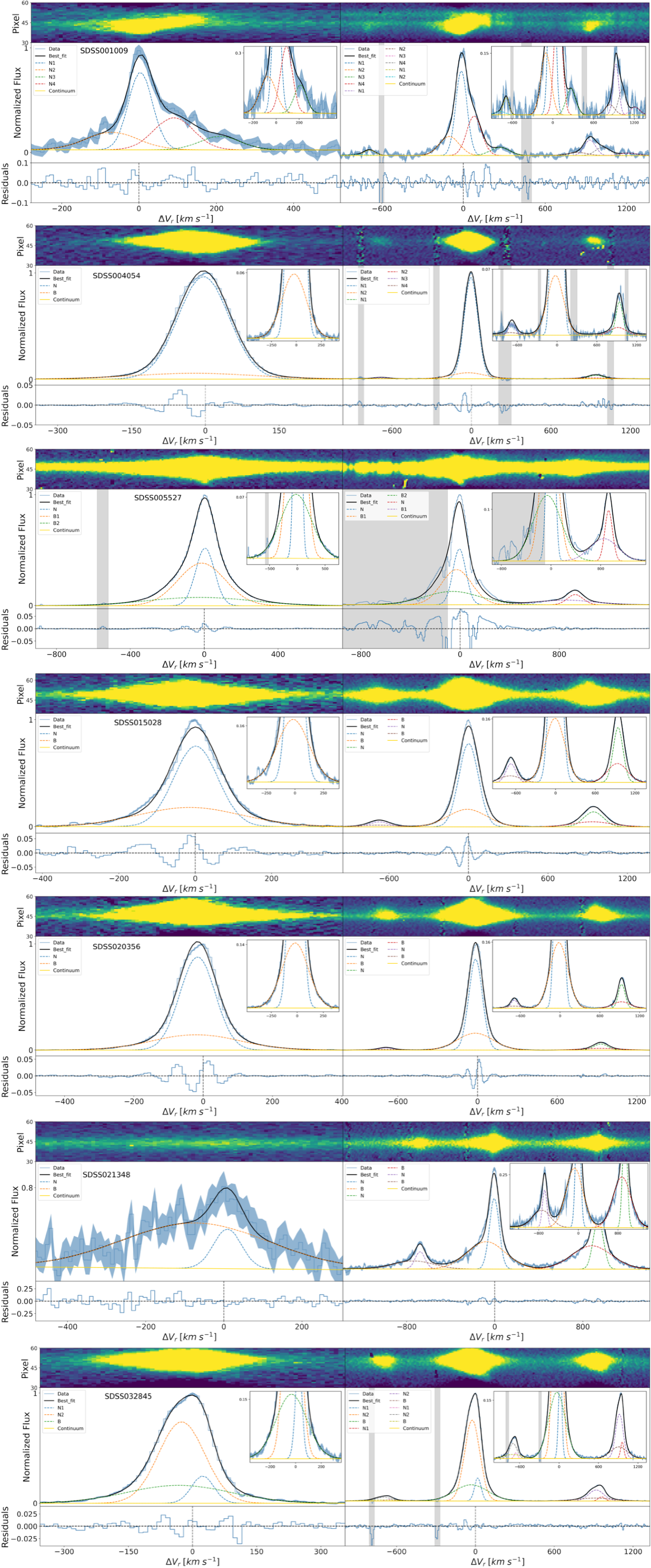}
    \end{minipage}
    \begin{minipage}{0.49\textwidth}
        \includegraphics[width=\linewidth]{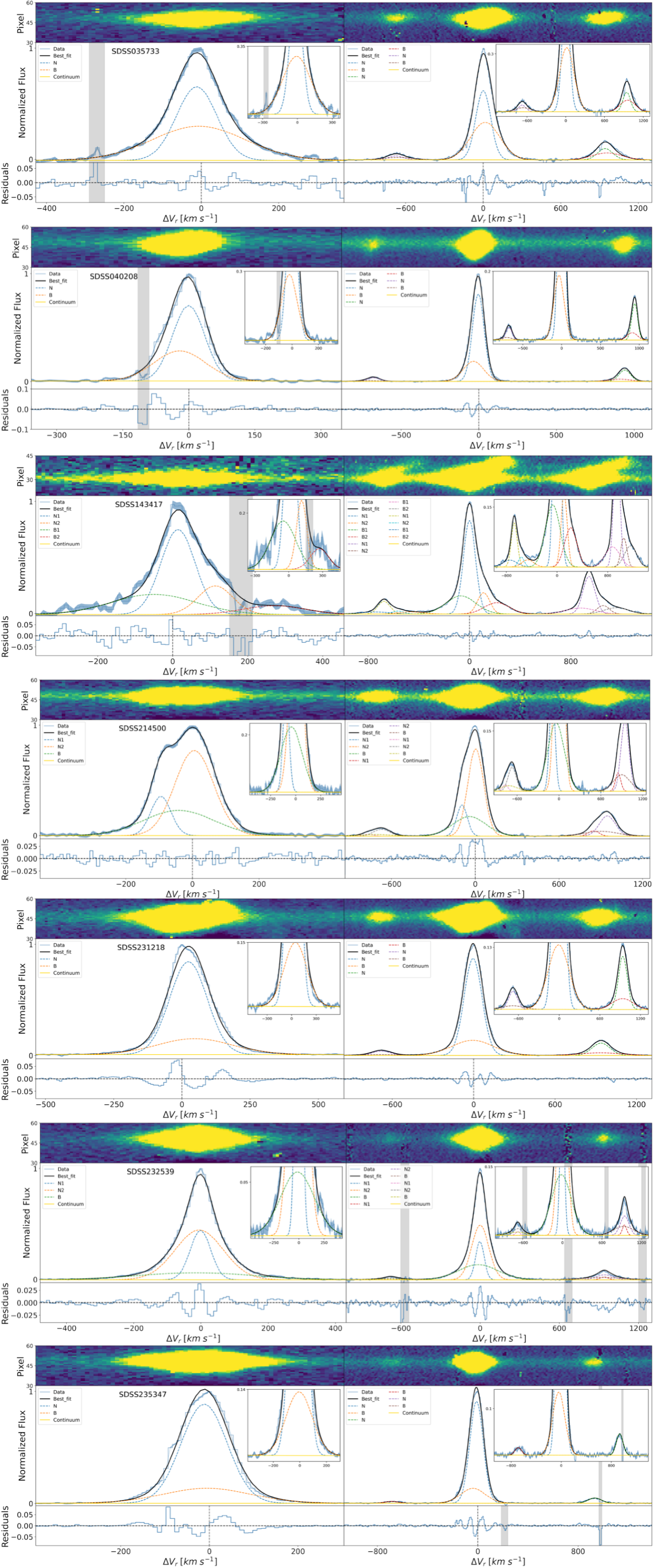}
    \end{minipage}
    \caption{Multi-Gaussian fitting of bright emission lines for the galaxies in our sample. The top panels show two-dimensional spectra, with the y-axis in pixel units. Center panels show Gaussian fits of [O{\sc iii}]$\lambda$5007$\AA$ (Left) and H$\alpha$ (Right). Bottom panels show fit residuals. The flux axis is normalized to the peak emission of each line. Spectra are shown in light blue (Data). The blue shadow represents the variance spectrum. The black line models the fit. The dashed lines show the different fitted components. Grey shadows are flagged regions excluded from fits. The yellow line represents the continuum. The zoom-in insets for the faint line wings are included in the upper-right corner of each plot. }
    \label{fig:figura_completa}
\end{figure*}

\section{Results}\label{Result}
The results of the multi-component line fitting and the kinematic properties obtained for the entire sample from these models are presented in Table~\ref {summary_kinematic_1} and Appendix  \ref{appendix:emission-lines}. 
In Figures  \ref{fig:figura_completa} and \ref{fig:figura_completa_faint}, we present the multi-Gaussian fits for the galaxy sample.

The intrinsic velocity dispersion of the emission line ($\sigma$) is obtained from the observed velocity dispersion ($\sigma_{obs}$), after subtraction of the instrumental ($\sigma_{ins}$) and thermal ($\sigma_{ther}$) broadening as follows, 
\begin{equation}
    \sigma= \sqrt{\sigma_{obs}^2-\sigma_{ins}^2-\sigma_{ther}^2}
\end{equation}
where the $\sigma_{ins}$ corresponds to a nominal value of $\sigma_{ins}$ $\sim$14.3 $km,s^{-1}$ for the X-Shooter spectrograph in the visual arm. 
The thermal broadening, caused by random thermal motions of the gas, is described by,
\begin{equation}
    \sigma_{ther}=\frac{ c \cdot \lambda_0}{\lambda}\cdot \sqrt{\frac{k_B\cdot T_{e}}{m_{ion}\cdot c^2}}
    \label{sigma_termico}
\end{equation}
where $k_{B}$ is the Boltzmann constant, $m_{ion}$ is the ion mass, and $T_{e}$ is the electron temperature. In our sample, six galaxies present detections of the auroral line [O\,\textsc{iii}] $\lambda 4364$, with electron temperatures determined by \cite{Loaiza} ranging from 9.400 to 11.900~K, with an average of $\approx 10.400$~K. Therefore, for galaxies with no auroral line detections, we assume $T_e = 10.000$~K, which is consistent with the measured values and with the typical temperatures expected for H\,\textsc{ii} regions in star-forming galaxies.

\begin{table*}
    \centering
        \caption{Summary of the kinematic fitting results.}
        \label{summary_kinematic_1}
        \setlength\tabcolsep{4pt}
        \small
        \begin{tabular}{lcccc|lcccc}
    \hline
 Ion Comp.$^{c}$  & $\sigma_{int}$ $^{d}$ & $\Delta$ $V_r$ $^{e}$ & Flux $^{f}$     & EM $^{g}$ &  Ion Comp.$^{c}$  & $\sigma_{int}$ $^{d}$ & $\Delta$ $V_r$ $^{e}$ & Flux $^{f}$     & EM $^{g}$ \\ \hline
                  & SDSS001009         &  $\chi^2_\nu $ $^{b}$$=$1.7, 1.7        &           &                   & SDSS143417                &  $\chi^2_\nu $ $^{b}$$=$3.6, 1.2        &          \\ \hline    
 H$\alpha$  N1    & 38.3$\pm$2.7       & -6.5 $\pm$2.00     & 29.9$\pm$6.0          & 46.1      & H$\alpha$  B1     & 114.9$\pm$7.1      & -85.2 $\pm$17.8    & 192.0$\pm$11.8          &  24.1     \\
            N2    & 71.5 $\pm$21.3     & -95.2 $\pm$44.0    & 15.3$\pm$6.4          & 18.2      &            B2     & 95.3$\pm$5.9      &  198.0$\pm$12.7    & 101.0$\pm$7.8          & 12.7      \\
            N3    & 68.5$\pm$10.0      & 275.5 $\pm$0.0     & 4.7$\pm$0.6           & 7.4       &            N1     &  44.6$\pm$0.6      & -17.7 $\pm$0.3    & 410.0$\pm$16.0          &  51.5      \\
            N4    & 52.5$\pm$2.7       & 87.8 $\pm$0.00     & 18.0$\pm$0.9          & 28.3      &            N2     & 44.6$\pm$2.5      &  93.7$\pm$0.0    & $93.6\pm$9.4          & 11.8      \\\hline

[O{\sc iii}]$_{5007}$ N1   & 27.8$\pm$2.3       & -4.7 $\pm$2.3      & 17.2$\pm$1.5 & 40.2      &  [O{\sc iii}]$_{5007}$ B1   & 110.7$\pm$0.0       & -86.0 $\pm$24.0      & 38.3$\pm$8.1 & 27.4    \\
             N2   & 71.7 $\pm$15.5     & -78.6 $\pm$0.0     & 8.9$\pm$1.7           & 20.6      &               B2   & 80.9$\pm$15.1      &  235.7$\pm$0.0     & 12.1$\pm$2.8          & 8.7   \\
             N3   & 51.0$\pm$13.7      & 201.0 $\pm$0.0     & 5.1$\pm$1.4           & 11.9      &               N1   & 43.3$\pm$13.5      &  -21.4$\pm$1.6     & 66.5$\pm$6.1          &  47.7  \\
             N4   & 48.9$\pm$9.6       & 81.3 $\pm$0.0      & 11.7$\pm$2.1          & 27.3      &               N2   & 43.8$\pm$5.5      &  78.3$\pm$0.0     & 22.5$\pm$2.5          & 16.2   \\ \hline
                  & SDSS004054\,      & $\chi^2_\nu $ $^{b}$ $=$ 1.74, 2.7         &           &                    &  SDSS214500     & $\chi^2_\nu $ $^{b}$ $=$ 2.7, 0.8         &         \\ \hline
 H$\alpha$  B     & 121.7$\pm$4.8      & -20.6 $\pm$2.9     & 60.6$\pm$5.5          & 12.6      &   H$\alpha$  B1     & 117.2$\pm$ 3.1     & -8.8$\pm$2.5      & 172.2$\pm$15.3         & 27.3   \\
            N     & 49.2$\pm$0.2       & 22.82 $\pm$0.2     & 420.2$\pm$3.8         & 87.4      &              N1     & 33.9$\pm$1.3      & -61.9$\pm$1.6      & 87.7$\pm$8.9         & 13.9   \\ 
                &       &      &          &      &              N2     & 54.7$\pm$0.9      & 36.7$\pm$1.1      & 370.6$\pm$6.8         &58.8    \\ \hline

[O{\sc iii}]$_{5007}$ B  & 120.9$\pm$6.0 & -18.5$\pm$3.6      & 113.2$\pm$14.1      & 12.7      &  [O{\sc iii}]$_{5007}$ B1  & 103.6$\pm$5.9 & -13.1$\pm$6.9       & 63.3$\pm$10.7       & 30.1    \\
             N    & 49.9$\pm$0.3       & 25.1$\pm$0.2       & 822.5$\pm$9.8         & 87.3      &               N1    & 26.5$\pm$1.5        & -67.1$\pm$1.4       & 28.5$\pm$3.1       &13.5   
              \\ 
                &       &      &          &      &              N2     & 56.3$\pm$1.8      & 32.9$\pm$1.5      & 118.7$\pm$7.2         &56.4    \\ \hline
                  & SDSS005527      & $\chi^2_\nu $ $^{b}$ $=$ 158.2, 2.5 &      &           &                    & SDSS231812\,       & $\chi^2_\nu $ $^{b}$ $=$2.3, 5.5  &      &         \\ \hline
 H$\alpha$  B1    & 223.9$\pm$10.8     & -27.9$\pm$5.0      & 741.7$\pm$36.7        & 29.2      &   H$\alpha$  B    &  152.8$\pm$2.1       & -2.3$\pm$0.8      &  174.1$\pm$6.4     &26.5     \\
            B2    & 120.1$\pm$5.9      & 5.6$\pm$2.1        & 1064.4$\pm$48.1       & 41.9      &              N    &  63.7$\pm$0.3       & 11.6$\pm$0.2      &  483.6$\pm$4.4     & 73.5    \\
            N     & 49.9$\pm$1.6       & 28.0$\pm$0.8       & 733.2$\pm$27.9        & 28.9      &                   &        &       &       &     \\\hline
[O{\sc iii}]$_{5007}$ B1 & 274.7$\pm$5.6 & 5.3$\pm$2.5        & 674.4$\pm$42.5      & 21.6      &  [O{\sc iii}]$_{5007}$ B & 136.5$\pm$7.5   & 22.8$\pm$4.3    &  140.1$\pm$2.5      & 25.4   \\
             B2   & 127.0$\pm$1.6      & 15.5$\pm$0.6       & 1585.6$\pm$22.8       & 50.9      &               N   &        67.1$\pm$1.3   & 5.6$\pm$0.7    &  410.9$\pm$17.7      &  75.6  \\
             N    & 49.0$\pm$0.5       & 33.7$\pm$0.3       & 858.6 $\pm$18.2       & 27.5      &                   &      &     &        &    \\ \hline
                  & SDSS015028      & $\chi^2_\nu $ $^{b}$ $=$ 4.5, 2.2  &       &           &                    & SDSS232539      & $\chi^2_\nu $ $^{b}$ $=$ 1.2, 1.7  &       &         \\ \hline
 H$\alpha$  B     & 146.8$\pm$2.5      & -9.4$\pm$ 1.0      & 380.6$\pm$22.4        & 29.7      &   H$\alpha$  B1     &    148.6$\pm$3.6     & -33.4$\pm$ 1.6      & 91.9$\pm$6.6   &  32.0  \\
            N     & 71.9$\pm$0.5       & 7.0$\pm$0.3        & 900.8$\pm$16.0        & 70.3      &              N1     &    27.4$\pm$1.5     & -20.5$\pm$0.5       & 50.4$\pm$7.2   & 17.6  \\
                 &       &       &       &      &              N2     &   62.4 $\pm$2.1     & -18.2$\pm$0.5       & 144.8$\pm$8.7   & 50.4   \\\hline
[O{\sc iii}]$_{5007}$ B  & 123.5$\pm$3.7 & -13.6$\pm$1.7      & 328.3$\pm$26.1      & 45.0      &  [O{\sc iii}]$_{5007}$ B1  & 165.5$\pm$12.8 & -30.3$\pm$5.9      & 70.4$\pm$14.8       &18.6    \\
             N    & 51.6 $\pm$1.1      & 12.2$\pm$0.7       & 402.0$\pm$17.8        & 55.0      &               N1    &  26.8$\pm$1.0     &  -17.2$\pm$0.5        & 92.4$\pm$7.3      & 24.3   \\
                 &       &       &       &      &              N2     &   69.6 $\pm$2.5     & -21.1$\pm$0.8       & 216.8$\pm$8.8   & 57.1   \\\hline
                  & SDSS020356      & $\chi^2_\nu $ $^{b}$ $=$ 3.3, 3.0     &    &           &                    & SDSS235347     & $\chi^2_\nu $ $^{b}$ $=$ 2.4, 6.8   &    &         \\ \hline
 H$\alpha$  B     & 131.9$\pm$1.2      & -2.7$\pm$ 0.5      & 332.1$\pm$7.4         & 32.7      &   H$\alpha$  B     &  95.1$\pm$3.7     & -28.0$\pm$3.2      & 73.9$\pm$11.5         &  20.7   \\
            N     & 48.7$\pm$0.2       & 61.2$\pm$0.1       & 685.0$\pm$4.6         & 67.3      &              N     &  50.9$\pm$0.6     & 21.4$\pm$0.4      & 284.0$\pm$8.9         & 79.3    \\\hline
[O{\sc iii}]$_{5007}$ B  & 125.4$\pm$3.3 & -4.0$\pm$1.6       & 346.9$\pm$22.2      & 29.0      &  [O{\sc iii}]$_{5007}$ B  & 97.5$\pm$8.2 &   6.0$\pm$0.5      & 160.4$\pm$33.7      & 27.5    \\
            N     & 49.1$\pm$0.5       & 64.2$\pm$0.3       & 849.8$\pm$14.9        & 71.0      &              N     & 49.8$\pm$1.1        & 21.4$\pm$0.5       & 422.0$\pm$24.4       & 72.5    \\ \hline
                  & SDSS021348      & $\chi^2_\nu $ $^{b}$ $=$ 1.4, 0.6   &      &           &                    & SDSS035733      & $\chi^2_\nu $ $^{b}$ $=$1.0, 2.9  &      &         \\ \hline
 H$\alpha$  B     & 170.9$\pm$2.9      & -72.8$\pm$2.6      & 78.1$\pm$1.9          & 59.9      &   H$\alpha$  B     & 111.4$\pm$2.4        &  15.1$\pm$0.7       & 186.1$\pm$12.7      &  51.3   \\
            N     & 41.1$\pm$0.7       & 2.4$\pm$0.6        & 52.4$\pm$1.1          & 40.1      &              N     & 55.4$\pm$1.1        &  -21.1$\pm$0.6       & 176.8$\pm$9.4      & 48.3    \\\hline
[O{\sc iii}]$_{5007}$ B & 187.5$\pm$13.8 & -82.3$\pm$8.0     & 17.5$\pm$1.5         & 85.8      &  [O{\sc iii}]$_{5007}$ B &  119.3$\pm$4.8 & 5.3$\pm$1.7     & 72.9$\pm$7.3         & 51.6    \\
             N    & 32.5$\pm$7.3       & 11.7$\pm$6.2      & 2.9$\pm$0.7            & 14.2      &               N    & 50.2$\pm$1.6        & -20.3$\pm$0.8     & 68.4$\pm$4.9         & 48.4    \\ \hline
                  & SDSS032845      & $\chi^2_\nu $ $^{b}$ $=$ 1.3, 1.1   &      &           &                    & SDSS040208       & $\chi^2_\nu $ $^{b}$ $=$2.2, 3  &      &         \\ \hline
 H$\alpha$  B    & 125.8$\pm$1.6      & -29.6$\pm$ 1.0     & 181.5$\pm$6.9         & 28.3      &   H$\alpha$  B    &  63.1$\pm$1.5   & -31.0$\pm$2.8     & 73.5$\pm$7.7   &  28.4   \\
            N1    & 55.3$\pm$0.4       & -21.3$\pm$0.4      & 409.4$\pm$5.9         & 63.8      &              N    &  34.1$\pm$0.5   & -17.2$\pm$0.4     & 185.4$\pm$6.7   & 71.6    \\
            N2    & 18.3$\pm$0.7       & 20.9$\pm$0.5       & 50.4$\pm$2.5          & 7.9       &         &     &  &    &     \\\hline
[O{\sc iii}]$_{5007}$ B & 119.4$\pm$5.1 & -28.0$\pm$3.0      & 127.7$\pm$16.5      & 29.7      &  [O{\sc iii}]$_{5007}$ B &  56.5$\pm$4.6 &  -20.3$\pm$5.5       & 76.2$\pm$31.0     & 39.0  \\
             N1   & 52.0$\pm$1.8       & -21.2$\pm$1.8      & 260.6 $\pm$15.9       & 60.6      &               N   &  33.4$\pm$2.2        &  -13.5$\pm$1.6       & 119.0$\pm$25.6     &  61.0   \\
             N2   & 21.5$\pm$1.4       & 28.0$\pm$1.2       & 42.0$\pm$4.6          & 9.7       &         &         &        &    &     \\ \hline
    \end{tabular}
    \normalsize
    \begin{tablenotes}
{\item (a) Rest-frame wavelength at $\r{A}$; (b) Reduced $\chi^2$ for \ha\ and [\oiii], respectively; (c) Broad (B) and Narrow (N) components; (d) Intrinsic velocity dispersion in km\,$s^{-1}$; (e) Radial velocity in km\,$s^{-1}$; (f) Flux of the component in units of $10^{-17}$ erg\,s$^{-1}$\,cm$^{-2}$; (g) Emission measure relative to the total flux of the line.}
    \end{tablenotes}
    \label{results_1}
\end{table*}

\subsection{Emission-line kinematics}

All VIS spectra show complex emission-line profiles. Single Gaussians are insufficient, as asymmetries and broad wings leave significant residuals.
We used two to four Gaussian components, depending on line shape and galaxy (see Table~\ref{summary_kinematic_1}). Eight galaxies required two components, four needed three, and two demanded four to reproduce their complex profiles. Following \citet{Bosch2019}, the need for additional components was assessed using the $\Delta(AIC)>$\,10 criterion.

Emission lines were modeled with narrow ($\sigma<$\,90\,km\,s$^{-1}$) and broad ($\sigma>$\,90\;km\,s$^{-1}$) components, both resolved.  
This division reflects expected rotation velocities: a $\sim$5$\times$10$^{9}$\,M$_ {\odot}$ galaxy rotates at $\sim$90\;km\,s$^{-1}$ \citep{Simons_2015}. Thus, if narrow components trace disk gas, their $\sigma$ should not exceed that value \citep[see also][]{Hogarth}.

In Section~\ref{l_sigma}, we further examine this assumption using the $L$–$\sigma$ scaling relation from \citet{Terlevich_1981}. Regarding the emission measure (EM), defined as the fraction of flux contributed by each component relative to the total line flux, we find that, in most cases, the narrow component dominates, while the broad component contributes 10–40\% of the total emission.

[O{\sc iii}]$\lambda$5007$\AA$ and H$\alpha$ show consistent radial velocities and velocity dispersions for both components, though fitted independently. This is illustrated in Fig.\ref{fig:sigma_comparison}, which presents a comparison between $\sigma_{\rm  H\alpha}$ and $\sigma_{\rm [O{III}]}$ for our sample.
While such an agreement may not necessarily be required and depends on the physical conditions and location of the gas that each component reflects, our result demonstrates robustness in determining the kinematics of ionized gas in these galaxies and suggests that recombination lines (H$\alpha$) and forbidden lines of high-ionization metal ions ([O{\sc iii}]) trace gas in similar regions. 

Narrow-component dispersions agree with spatially resolved results from \citet{Goncalves_2010} for a subset of LBAs. Those authors found SFR-$\sigma$ correlations interpreted as star formation-driven turbulence. Intense star formation injects energy into the ISM, broadening lines slightly beyond virial expectations. 
As we show later in Sect.~\ref{l_sigma}, our results are also consistent with an increasing velocity dispersion for galaxies with stronger star formation. High-resolution IFU studies are required to disentangle rotation, turbulence, and outflows in LBAs \citep[e.g.][]{Bik2018,Arroyo-Polonio2024}.

\subsection{The luminosity of narrow and broad kinematic components}
\label{l_sigma}

All bright lines require at least one narrow and one broad component. The narrow component traces bright star-forming regions and contributes $>$\,80\% of total flux.
Their physical properties, such as dust extinction and density, show significant differences with those of the broader components, suggesting that the underlying physical mechanisms of each component may also differ. 

To demonstrate this difference, we analyze the well-known $L-\sigma$ correlation \citep{Terlevich_1981,Chavez_2014}, where $L$ and $\sigma$ represent the $H\beta$ luminosity and intrinsic velocity dispersion, respectively. 
Giant H{\sc ii} regions and H{\sc ii} galaxies (compact low-mass starbursts) follow an $L-\sigma$ relation, a robust cosmological distance indicator \citep{Terlevich_1981,Chavez_2014, FernandezArenas2018,Chavez2025}. 
The underlying assumption for this correlation, which has been shown to work for low and high-redshift compact starbursts \citep[e.g.][]{AnaLuisa2021,Chavez2025}, is that the narrow components of the emission lines are usually interpreted as virial motions of gas gravitationally bound to massive star clusters. However, high-resolution IFU observations of local analogs show that star formation is clumpy, with relative motions among knots that can also affect the narrow-line kinematics \citep{Santos_junior_2025}.  Thus, the observed narrow components may reflect both virialized gas within clumps and relative clump-to-clump motions \citep[see also][]{Amorin_2012b}.

\begin{figure}
    \centering
    \includegraphics[scale=0.48]{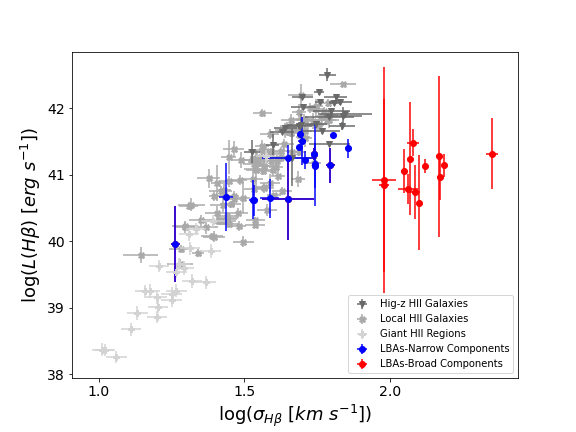}
    \caption{L-$\sigma$ relationship. The blue and red circles correspond to the narrow and broad components of our sample of LBA galaxies, respectively. Grey symbols are Giant HII regions and HII galaxies at $0 < z< 2.33$ from \cite{Terlevich2015}. 
    }
    \label{fig:relacion_L-sigma}
\end{figure}

In Figure \ref{fig:relacion_L-sigma} we show the  L-$\sigma$ relation for 156 low and high-redshift  ($0 \lesssim z \lesssim 2.3$) starbursts from \citet{Terlevich2015}. We include the values for our sample, considering the individual, narrow (blue), and broad (red) components. 
To determine the luminosities for each component, we use the extinction-corrected $H\beta$ flux, adopting the  \cite{Calzetti_2001} extinction law.
The narrow components fitted to our sample follow the $L-\sigma$ relation of H{\sc ii} galaxies, with $\sigma_{H\beta}$ values up to $\sim$ 60 km,s$^{-1}$. Some galaxies exhibit $\sigma_{H\beta}$ slightly larger for their $H\beta$ luminosities compared to extragalactic \hii\ regions. 
Ionized gas bound to massive clusters may also include clump-to-clump relative motions contributing to narrow-line kinematics. The broad components, instead, show significantly larger $\sigma_{H\beta}$ than expected for their $H\beta$ luminosity if they were purely virialized, indicating the presence of an additional mechanism driving gas turbulence.

\subsection{The nature of the broad emission line components}

Several mechanisms may cause line broadening in compact starbursts.
These mechanisms include stellar winds from massive stars, expansion of supernova remnants (SNe), outflows driven by SNe or turbulent mixing layers (TML), and the presence of an AGN \citep[e.g.][]{Izotov_2007, Amorin_2012b, Hogarth, Martin2024}. Studying a sample of nearby Blue Compact Dwarf galaxies (BCDs) with broad [\oiii] and \ha\ emission,  \cite{Izotov_2007} derive the observed  H$\alpha$ luminosity of the broad components ($10^{36}$-$10^{39}$ erg $s^{-1}$), which are expected from the interaction between dense circumstellar envelopes around hot stars with stellar winds and/or supernova remnants. The LBAs, instead, show H$\alpha$ luminosities between $10^{40}$-$10^{42}$ erg $s^{-1}$ that are $\sim$2 dex larger than that of BCDs (Fig.~\ref{fig:relacion_L-sigma}). Such a high broad line luminosity is more typically associated with Type II SNe or an AGN. 

An AGN origin is disfavored by the absence of clear non-thermal emission.
Broad components in both [O{\sc iii}]$\lambda$5007$\AA$ and H$\alpha$ share similar kinematics, further arguing against an AGN origin.Although an AGN cannot be ruled out, confirmation would require deep X-ray or high-resolution radio data to detect any non-thermal source. Several LBAs have already been observed in X-rays; for instance, \citet{Basu-Zych2013} reported elevated 2–10 keV luminosities relative to the SFR, consistent with enhanced high-mass X-ray binary (HMXB) populations rather than AGN activity. 
Previous works by \citet{Jia_2011} using x-ray data and \citet{Alexandroff_2012} using radio continuum observations highlight the complex challenge of confirming AGN activity and quantifying its contribution to the observed galaxy properties. 
In the absence of strong non-thermal signatures, the broad components are most likely powered by stellar feedback.

Turbulent mixing layers (TMLs) between cold disk gas and hot outflows can also broaden lines \citep{Westmoquette_2008, Binette_2009}.
By implementing a set of TML models, \cite{Binette_2009} found broad components with considerably supersonic velocities (FWHM $\sim$ 2400 km $s^{-1}$).  In galaxies, these high-velocity components are rarely observed in bright emission lines like [O{\sc iii}]$\lambda\lambda$4959,5007$\AA$, H$\alpha$, and H$\beta$, such as in Mrk 71, a very young giant HII region ($<3$Myr) belonging to the irregular dwarf galaxy NGC 2366 \citep{Komarova2021}, or in the giant H{\sc ii} region NGC\,5471 \citep{Castaneda1990}. 
However, TML models generally assume a high-ionization parameter $\log(U)$, leading the ionization throughout the turbulent layer, thus making the gas dominated by highly excited species and therefore
 struggling to explain such components in low-ionization forbidden lines, such as [N{\sc ii}] or [S{\sc ii}] \citep{Binette_2009}.
For example, \citet{Bosch2019} and \citet{Hogarth} discussed TMLs as one of the mechanisms to explain the complex kinematics of Green Pea galaxies showing broad components in [N{\sc ii}] and [S{\sc ii}], concluding that a TML model would require a low $\log(U)$ \citep{Binette_2009}. However, using the [O{\sc ii}]/[O{\sc iii}] ratio of GP\,1429, \cite{Hogarth} found a relatively high $\log(U)$ for the broad component, disfavoring the TML interpretation.

In the LBAs, we do not find broad components reaching such high velocities of $>$1000\kms. 
LBAs show no $>$1000\kms\ broad components and display relatively high $\log(U)$ ($–2.9$ to $–2.3$ \citet{Loaiza}). This, along with the relatively modest velocities found by our fittings of both low- and high-ionization lines, suggests that the broadening of the faint line wings cannot be entirely explained by TMLs. TML contributions cannot be excluded, but detailed modeling and high-dispersion IFU mapping are needed to test them.

Finally, shocks produced by young supernova remnants can significantly contribute to the turbulence of ionized gas. According to \cite{ho_2014}, the contributions of shock processes can be identified in diagnostic diagrams by the presence of broad components with higher [N{\sc ii}], [S{\sc ii}], [O{\sc ii}] to H$\alpha$ ratios, as well as elevated electron density values. Therefore, a scenario with a strong contribution from stellar feedback and shocks from supernova remnants is a likely mechanism for providing energy to the turbulent ionized gas in these star-forming galaxies, resulting in high velocity dispersions in the broad components. 
Given the low broad-to-narrow luminosity ratios and lack of AGN signatures, stellar winds and supernova feedback likely power the broad emission.

\subsection{Emission-line diagnostics}

\begin{figure}[t!]
    \centering
    \includegraphics[scale=0.45]{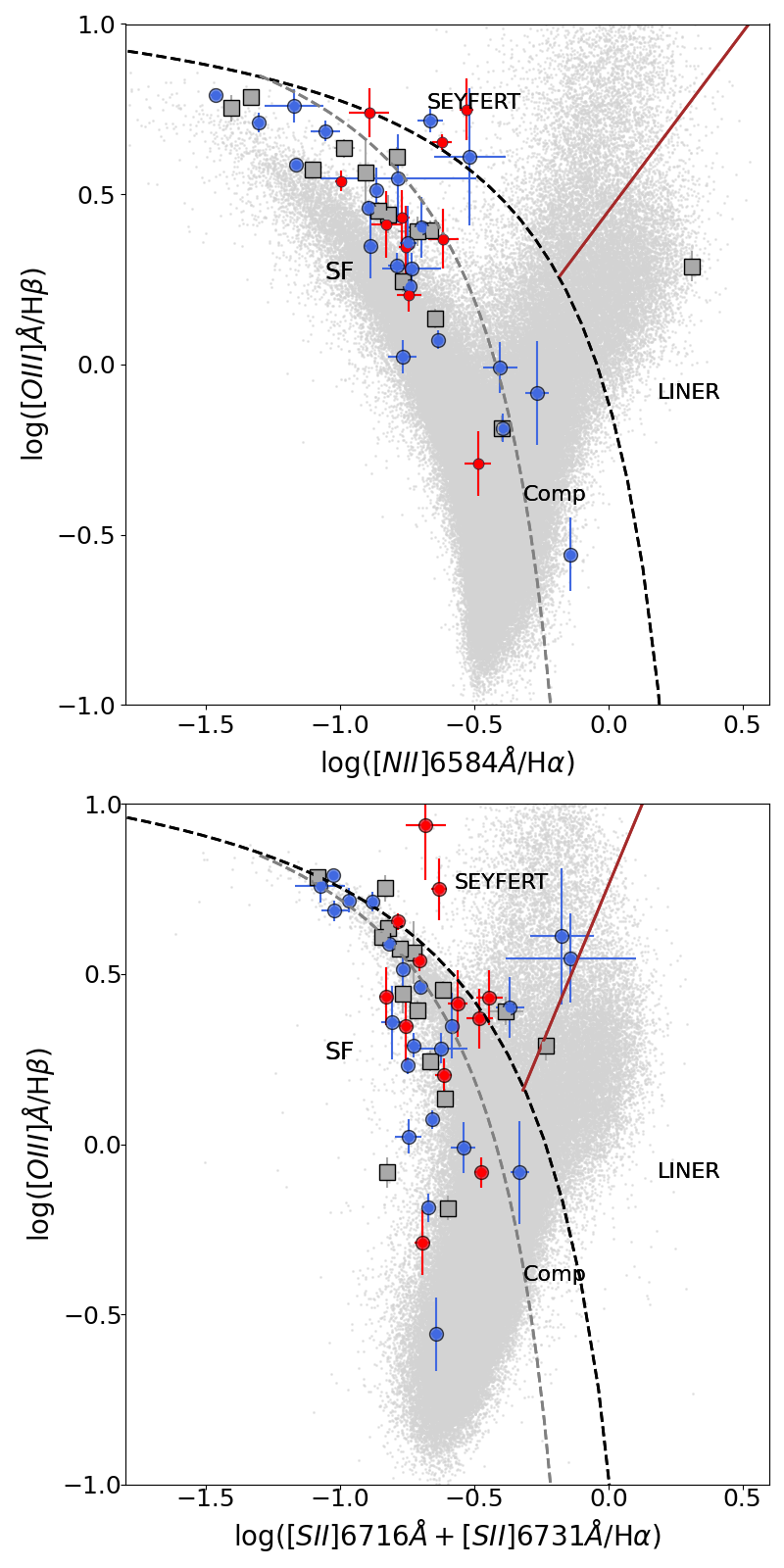}
    \caption[BPT]{Classic  diagnostic diagrams based on emission line ratios \citep{Baldwin, Veilleux_1987} [O{\sc iii}]$\lambda$5007/$H\beta$ vs. [N{\sc ii}]$\lambda6584/H\alpha$  (Top) and [O{\sc iii}]$\lambda$5007/ $H\beta$ vs. [S{\sc ii}]$\lambda \lambda $6716.6731/$H\alpha$ (Bottom). Symbols show emission lines integrated fluxes (gray squares),  narrow (blue circles) and broad (red circles) Gaussian components for each galaxy.  
    The gray dots correspond to the SDSS-DR7 MPA-JHU galaxy sample from \citet{Perez-Montero2021}. Regions of different excitation mechanisms are labeled and established by theoretical and empirical demarcation lines of \cite{Kewley_2001, Kewley_2006} (black dashed), \cite{Kauffmann_2003} (gray dashed), and \cite{Schawinski_2007} (brown solid). }
    \label{fig:bpt_simple}
\end{figure}

We examine classic emission-line diagnostic diagrams using our multi-Gaussian fits. Our goal is to assess excitation properties and the dominant ionization mechanism of each component.
We show in Fig.~\ref{fig:bpt_simple} the diagnostic diagrams based on the [O{\sc iii}]$\lambda$5007/$H\beta$ vs. [N{\sc ii}]$\lambda$6584/H$\alpha$ relation \citep{Baldwin} and [S{\sc ii}]$\lambda\lambda$6716,6731/$H\alpha$ \citep{Veilleux_1987}. 
In Fig.~\ref{fig:bpt_simple}, we show the ratios based on integrated line flux for each galaxy with gray squares. 
Blue and red circles show instead the ratios for the narrow ($\sigma \lesssim 90$\,\kms) and broad ($\sigma \gtrsim 90$\,\kms) components. For comparison, we show SDSS galaxies and the different theoretical and empirical demarcation lines, which allow us to constrain the ionization source for each component. 
Most galaxies are consistent with stellar photoionization; only SDSS021348 shows elevated N2 and S2 ratios typical of LINERs.

There is a small fraction of galaxies showing kinematic components within the so-called "composite" region, thus allowing a contribution of shocks or some nuclear activity. Three galaxies, SDSS001009, SDSS005527, and SDSS015028 have their kinematic components consistent with high ionization AGN Seyfert II excitation.  

Overall, no systematic trend appears among the three diagnostic ratios in Fig.~\ref{fig:bpt_simple}. In particular, the broad components are not associated with a different excitation mechanism compared to the narrow components, and both are consistent with the conclusions one can obtain from the integrated line ratios.

\subsection{Electron density}\label{sec:density}

Electron density ($n_{e}$) is key to characterizing physical conditions in star-forming regions, together with electron temperature.
We determined $n_{e}$ using the Sulfur doublet [S{\sc ii}]$\lambda\lambda$6716/6731$\r{A}$, because it is present in the entire sample and because the [O{\sc ii}]$\lambda\lambda$3726/3729$\r{A}$ doublet is observed in the UBV arm of X-Shooter and therefore has a factor of $\sim$1.65 lower in spectral resolution. 
We computed $n_e$ using Specsy \citep{Fernandez_2023}.  
The electron temperatures for 6 galaxies (SDSS004054, SDSS005527, SDSS015028, SDSS020356, SDSS032845, SDSS235347) were taken from \cite{Loaiza}. 
For the remaining eight galaxies without $T_e$ measurements, we assumed $T_e =$\,10 000\,K.

Table \ref{tab:density} includes the density values obtained for the integrated flux ratios and for those of the different kinematic components following the above methodology. 
Electron densities range from 120 cm$^{-3}$ to 1416 cm$^{-3}$  with (narrow) and 366 cm$^{-3}$ (broad). For some galaxies (SDSS005527, SDSS020356, SDSS214500, SDSS231812), the broad components show higher density than the narrow components.  
Two galaxies (SDSS015028, SDSS0035733) show all the narrow, broad, and integrated components with very similar densities. Finally, three galaxies (SDSS021348, SDSS032845, SDSS232539) show denser narrow components compared to the broad components. 

While we acknowledge large uncertainties inherent to the electron density estimates that may affect the analysis, our results suggest that different kinematic components may have different density conditions according to their [\sii] ratios. 
These variations likely reflect the diversity in physical and dynamical conditions among LBAs. 
For example, effects such as the compactness, age, and ionization conditions of the bright star-forming regions may result in different gas pressure conditions for the narrow components tracing H{\sc ii} gas. Similarly, the presence of denser broad components tracing outflows can be the result of shocked gas due to SNe feedback and the interacting/merger nature of some of the galaxies, which appears consistent with our analysis based on classic emission-line (BPT) diagnostics \citep[e.g.][]{ho_2014, RodriguezDelPino2019}. 
For example, the electron density of the outflow components is consistent with those derived from stacked KMOS-3D spectra of galaxies at $z\sim0.6-2.7$, which are about 400 cm$^{-3}$ \citep{Forster-Schreiber2019}. 
Overall, the $n_e$ values of both components agree with those in other UV-bright starbursts and local high-z analogs \citep[e.g.][]{Arribas2014, Hogarth, Alvarez-Marquez2021, Amorin2024, Martin2024}.

We also compared our $n_e$ estimates with those reported by \citet{Loaiza} for the same LBA sample. Their analysis, based on single-component fits to the [\sii] doublet, yields average densities around 160\,cm$^{-3}$, whereas our global median values are somewhat higher, around 300\,cm$^{-3}$. This difference is likely due to our multi-component Gaussian fitting approach, which allows us to isolate and quantify denser kinematic components that contribute to the total emission. These results suggest that integrated single-component measurements may underestimate the density of the outflowing ionized gas when multiple components are present.

\subsection{Outflows properties}

From the L-$\sigma$ relation (Sect.~\ref{l_sigma}), the broad components trace gas that has been accelerated beyond typical H{\sc ii} region dynamics. These highly turbulent components are identified as ionized outflows, likely denser and more compact than those traced by UV absorption lines.
Its origin is probably a combination of gas driven by the combined action of strong stellar winds from massive stars, SNe explosions, and shocks originating from galaxy interactions.

We derived outflow properties from the broad components, following equations in \cite{Concas_2022} and \cite{llerena2023ionized}. 
Table \ref{tab:outflows_properties} shows a summary of our results.
We first focus on the outflow gas mass, which can be estimated as
\begin{equation}
\label{eq:mass_out}
    M_{out}^{H\alpha}= 3.2 \times 10^{5}\left ( \frac{L_{B}^{H\alpha}}{10^{40} \;erg\;s^{-1}} \right )\left ( \frac{100 cm^{-3}}{n_{e}} \right )\; M_{\odot}
\end{equation}
Where $L_{B}^{H\alpha}$ corresponds to the broadest component of the emission line of $H\alpha$, which was extinction-corrected using \cite{Calzetti_2001}. 
Following \citet{llerena2023ionized}, we use $H\alpha$ instead of [\oiii] since [\oiii]-based masses depend on poorly constrained metallicity effects. In Eq. \ref{eq:mass_out}, $n_{e}$ is the density of the broad component. In cases where the density of the broad component cannot be derived, we use the global density measured from the integrated line ratios (Table \ref{tab:density}).

Assuming a multi-conical or spherical geometry, the outflow mass rate ($\dot{M}_{out}$) is defined \cite{Lutz_2020}  as

\begin{equation} \label{eq:máss_out_rate}
\begin{split}
\dot{M}_{out}&=C \; \frac{M_{out}\;V_{out}}{R_{out}}
, \\
&= 1.02 \times 10^{-9}\left ( \frac{V_{out}}{km \;s^{-1}} \right )\left (\frac{M_{out}}{M_{\odot}}\right )\left ( \frac{kpc}{R_{out}} \right )\; C \;M_{\odot} \; yr^{-1},
\end{split}
\end{equation}
where C depends on the assumed outflow history, $R_{out}$ is the outflow radius, $M_{out}$ is the outflow mass and $V_{out}$ is the outflow velocity. 

\begin{table}[t!]
    \centering
    \begin{threeparttable}
    \caption{Outflow properties}
    \label{tab:outflows_properties}
    \small 
    \begin{tabular}{l|c|c|c|c}
    \hline
    ID & $\log (v_{\rm out})$ & $\log(\rm{SFR}_{\rm N})$ & $\log (\dot{M}_{\rm out})$ & $\log (\eta)$ \\
       & [km\,s$^{-1}$]       & [$M_{\odot}$\,yr$^{-1}$] & [$M_{\odot}$\,yr$^{-1}$]   & [dex] \\
    \hline
    001009 & -               & -               & -               & -               \\
    004054 & 2.42 $\pm$ 0.01 & 0.84 $\pm$ 0.01 & 0.06 $\pm$ 0.13 & -0.78 $\pm$ 0.14 \\
    005527 & 2.74 $\pm$ 0.01 & 1.04 $\pm$ 0.05 & 0.40 $\pm$ 0.10 & -0.64 $\pm$ 0.11 \\
    015028 & 2.42 $\pm$ 0.01 & 0.70 $\pm$ 0.04 & 0.43 $\pm$ 0.19 & -0.27 $\pm$ 0.19 \\
    020356 & 2.41 $\pm$ 0.01 & 0.64 $\pm$ 0.01 & 0.13 $\pm$ 0.03 & -0.77 $\pm$ 0.04 \\
    021348 & 1.80 $\pm$ 0.10 & -               & -               & -               \\
    032845 & 2.43 $\pm$ 0.02 & 0.57 $\pm$ 0.11 & -0.40 $\pm$ 0.22 & -0.97 $\pm$ 0.25 \\
    035733 & 2.37 $\pm$ 0.02 & 0.57 $\pm$ 0.07 & 0.33 $\pm$ 0.08 & -0.24 $\pm$ 0.10 \\
    040208 & 2.12 $\pm$ 0.04 & -               & -               & -               \\
    143417 & 2.30 $\pm$ 0.25 & 0.81 $\pm$ 0.05 & -0.69 $\pm$ 0.26 & -1.50 $\pm$ 0.26 \\
    214500 & 2.34 $\pm$ 0.03 & 0.67 $\pm$ 0.03 & 0.28 $\pm$ 0.13 & -0.39 $\pm$ 0.13 \\
    231812 & 2.40 $\pm$ 0.03 & 0.85 $\pm$ 0.01 & -0.31 $\pm$ 0.01 & -1.16 $\pm$ 0.05 \\
    232539 & 2.56 $\pm$ 0.03 & 0.50 $\pm$ 0.08 & 0.41 $\pm$ 0.10 & -0.09 $\pm$ 0.13 \\
    235347 & 2.28 $\pm$ 0.04 & 0.46 $\pm$ 0.04 & -0.05 $\pm$ 0.26 & -0.50 $\pm$ 0.26 \\
    \hline
    \end{tabular}
    \begin{tablenotes}
        \footnotesize
        \item (a) $\log (v_{\rm out})$: outflow velocity; (b) $\rm{SFR}_{\rm Narrow}$: star formation rate from the narrow component luminosity, no aperture corrections were applied; (c) $\dot{M}_{\rm out}$: mass outflow rate; (d) $\eta$: mass-loading factor ($\dot{M}_{\rm out}/\mathrm{SFR}$). 
    \end{tablenotes}
    \end{threeparttable}
\end{table}

Following  \cite{llerena2023ionized}, we adopt a constant C starting at $-t=-R_{out}/V_{out}$ which gives $C=$\,1 which corresponds to assuming that the wind mass rate is constant. To be consistent with previous works, we define the outflow velocity as $V_{out}=\mid \Delta_{V_r} - 2\times \sigma_{Broad} \mid$ \citep{Genzel_2011}, where we use the intrinsic velocity dispersion of H$\alpha$. As shown in Fig.~\ref{fig:sigma_comparison}, using instead [O{\sc iii}]$\lambda$5007$\AA$ has a negligible impact on the final results. For the outflow radius, we assume $R_{out}=R_{e}$ \citep{Forster-Schreiber2019} which is justified by the typical sizes of ionized gas outflows detected with high-resolution adaptive optics (AO)-assisted SINFONI observations for a sample of high-z galaxies (see \cite{Newman_2012} and \cite{Schreiber_2014}). 
Applying Eq.~\ref{eq:máss_out_rate} yields $\dot{M}_{out}=$\,0.20-2.72\,$M_{\odot}\;yr^{-1}$ ($<\dot{M}_{out}>=$\,1.43\,$M_{\odot}\;yr^{-1}$).

Finally, we obtain the mass loading factor $\eta$ as,
\begin{equation}\label{eq:máss_loading_factor}
    \eta=\frac{\dot{M}_{out}}{SFR_{narrow}}
\end{equation}
where $\dot{M}_{out}$ is obtained with Eq.~\ref{eq:máss_out_rate} and $SFR_{narrow}$ is determined using the luminosity of H$\alpha$, taking into account only the narrow component and applying the extinction correction described by \cite{Overzier_2009}. We note that our SFR values are systematically lower than those reported by \cite{Overzier_2009} for the same objects. This is expected, since they used SDSS spectra and applied an aperture correction factor of $\sim$1.7 to recover the total galaxy flux, while our estimates are based on VLT/X-Shooter data and refer only to the narrow component. In addition, the broad component contributes a smaller but non-negligible fraction to the total H$\alpha$ flux, partially accounting for this difference.

The mass-loading factor is a crucial parameter for understanding how outflows affect the properties of galaxies. This factor quantifies the efficiency with which outflows remove gas from the galaxy relative to the star formation rate.
The mass-loading factors span $\eta=$\,0.03–0.81 (mean $<\eta>=$\,0.31).

The values of outflow properties for our sample are summarized in Table \ref{tab:outflows_properties}. In Sect.~\ref{sec5}, we discuss these results and how they contribute to our understanding of starburst-driven feedback.

\section{Discussion}\label{sec5}

\subsection{Comparison with previous works}

Our sample of LBAs shows complex emission lines with both narrow and broad components. 
The velocity dispersion of the narrow components agrees with that reported by \citet{Goncalves_2010} from the central ($\sim$\,1$''$) Paschen~$\alpha$ emission of LBAs, who found relatively high mean values ($\sim$70\,km\,s$^{-1}$) indicative of turbulent gas kinematics. 
However, \citet{Goncalves_2010} did not detect clear high-velocity components, possibly due to the shallower depth of their NIR data.

For the nine galaxies overlapping with \citet{Goncalves_2010}, our kinematics agree closely. 
This suggests that highly turbulent ionized gas kinematics is common in LBAs, with values exceeding those of Giant HII regions in spiral galaxies \citep[e.g.][]{Firpo_2010} and similar to spectroscopically-selected starburst galaxies in the local universe \citep[e.g.][]{Ostlin2001, Green2010, Amorin_2012b, Chavez_2014}. 

The ionized gas emission of the narrow components follows the $L$–$\sigma$ relation shown in Fig.~\ref{fig:relacion_L-sigma} \citep[][see Sect.~\ref{l_sigma}]{Terlevich_1981}, implying gas gravitationally bound to massive star clusters. 
Additional sources of energy contributing to the observed turbulence may include stellar feedback and local enhancements, such as shocks driven by interactions or mergers \citep[e.g.][]{Baron2024, Martin2024}. 
This could explain some LBAs located in the outer envelope of that relation (Fig.~\ref{fig:relacion_L-sigma}), originally defined for young extragalactic H{\sc ii} regions.

Morphologically, the sample is diverse, including compact galaxies with a dominant central source and spatially extended systems showing tidal features \citep{Overzier_2009}. 
In several extended LBAs, the two-dimensional X-Shooter spectra are spatially resolved, revealing complex velocity fields. 
These characteristics resemble those of other local analogs to high-redshift galaxies, such as Haro~11 \citep{Ostlin2009, Bik2018} or the BCD galaxies studied by \citet{Bik_2022}. 
Overall, the complex ionized-gas kinematics of LBAs likely reflects a disturbed gas morphology, with strong gravitational instabilities affecting both stars and gas, probably driven by mergers or accretion events.

Their kinematics thus appear dominated by turbulence rather than ordered rotation, similar to low-mass starbursts at higher redshift \citep[e.g.][]{Wisnioski2015, Simons_2015}. 
Recently, \citet{deGraaff2024} analyzed the gas kinematics of star-forming galaxies at $5.5 < z < 7.4$ using JWST/NIRSpec R2700 spectra, finding comparable stellar masses, sizes, and velocity structures to LBAs, characterized by broadened lines, modest gradients, and mixed rotational–dispersion support with $\sigma \sim 30$–70\,km\,s$^{-1}$. 
These parallels reinforce the role of LBAs as nearby laboratories for studying the physical processes that shape galaxies in the early Universe.

\subsection{Trends for outflow rates and mass loading with stellar mass and compact star formation}

\begin{figure*}[t!]
    \centering
    \includegraphics[height=0.46\textwidth]{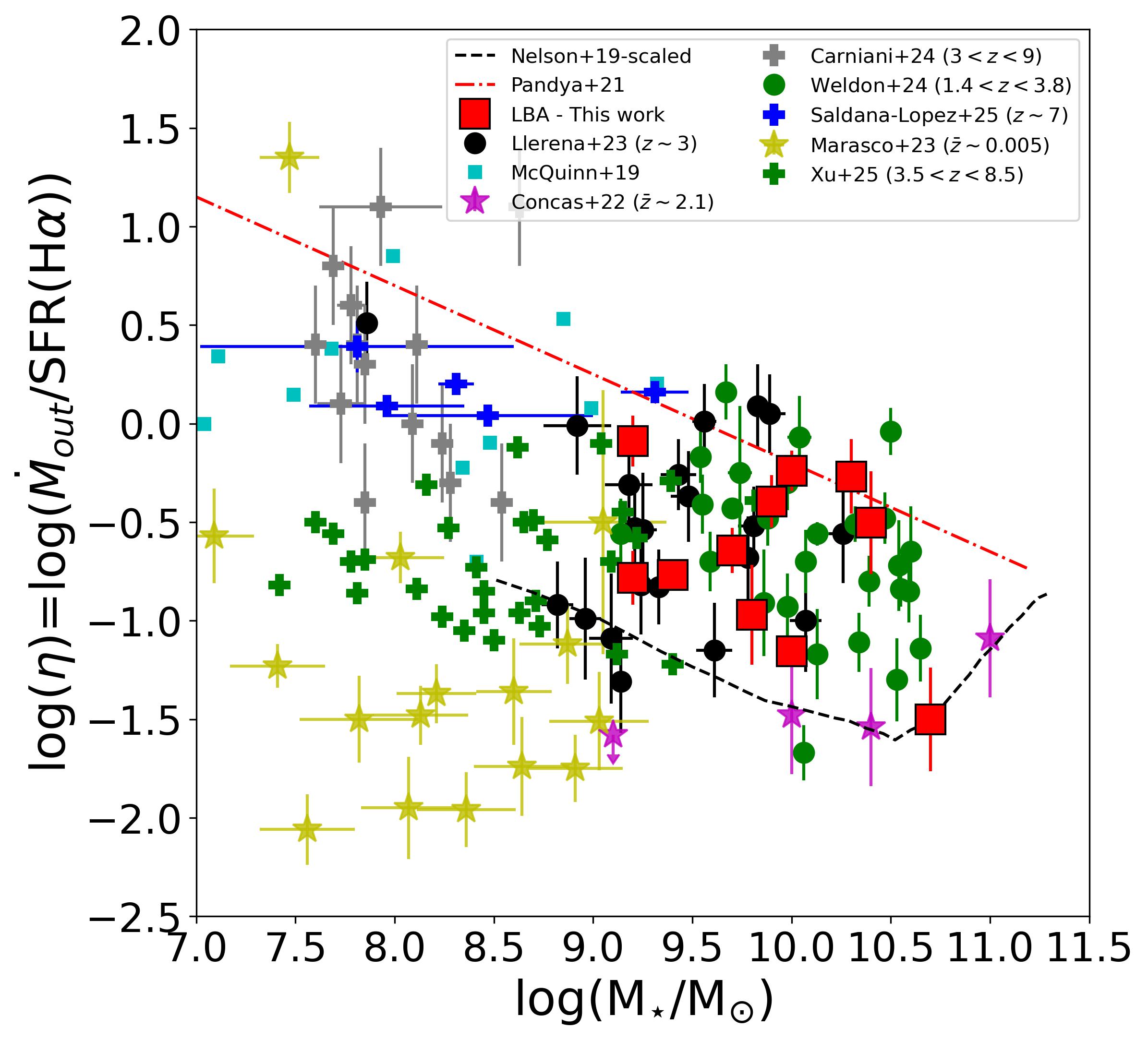}\label{fig:sub1}
    \hskip1ex
    \includegraphics[height=0.46\textwidth]{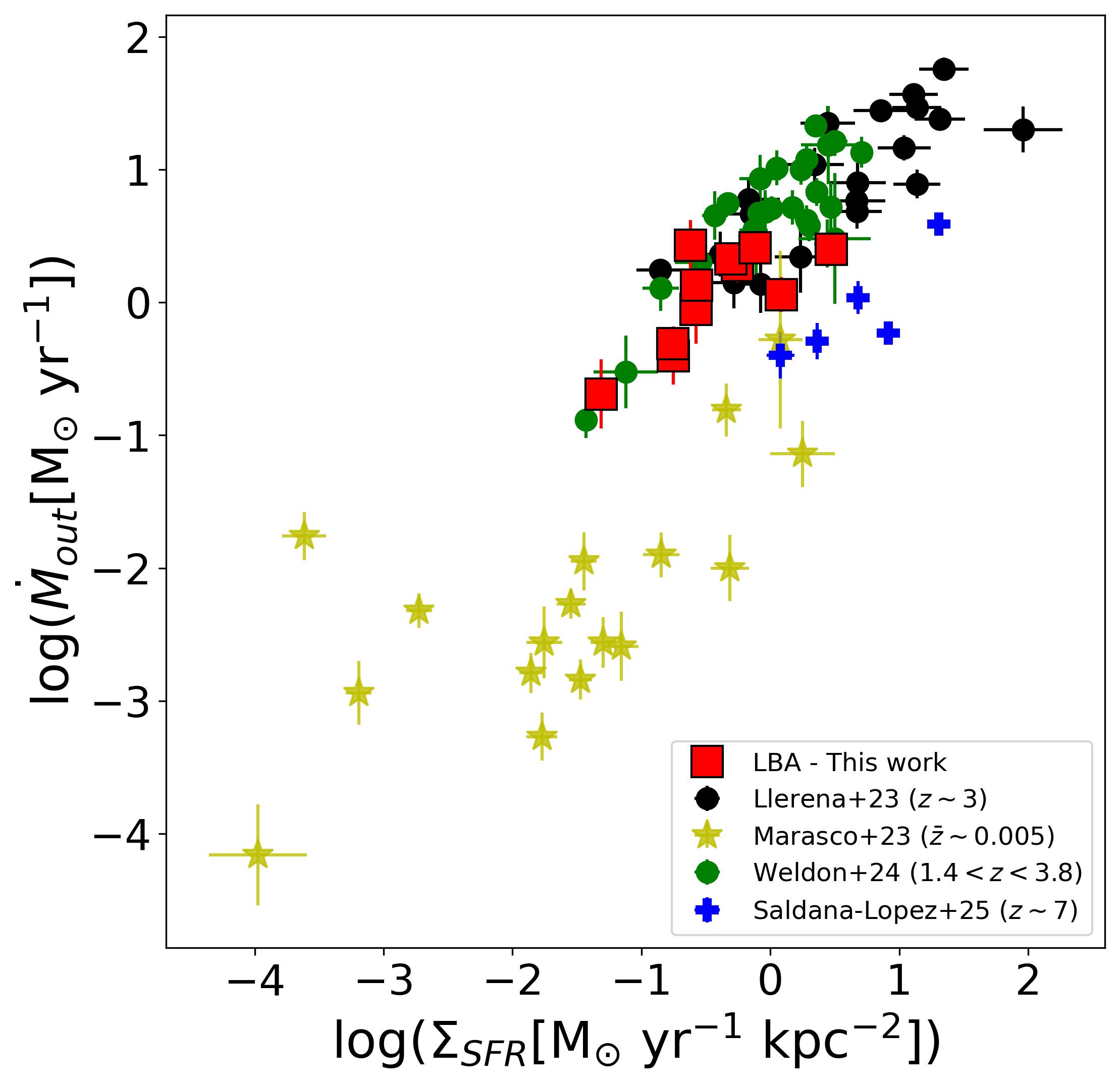}\label{fig:sub2}
    \caption{Outflow mass loading factor as a function of the stellar mass of galaxies (Left) and stellar mass outflow rate as a function of SFR surface density (Right). Our LBAs are shown in red squares. Green and black circles LBGs at $z\sim$\,1.4-3.8 from \citet{weldon2024} and \citet{llerena2023ionized}, respectively, while purple stars are mean-weighted averages from stacks of star-forming galaxies at z$\sim$ 2.1 of \cite{Concas_2022}. Gray and blue crosses are low-mass star-forming galaxies at $z\sim$\,3-9 observed with JWST/NIRSpec R=2700 spectroscopy from \cite{Carniani2024} and \cite{Saldana-Lopez2025}, respectively. The green crosses are the results from \cite{Xu2025} for galaxies at $z\sim3-9$. Yellow stars are local dwarf galaxies of \cite{Marasco_2023}. Light blue squares indicate local dwarf galaxies from \cite{Mcquin_2019}. Red dashed lines are rescaled relations from Ilustris-TNG simulations of \cite{Nelson_2019} and black dashed lines are from FIRE-2 simulations of \cite{Pandya_2021}.}
    \label{fig:eta_mass_Sigma}
\end{figure*}

The turbulent gas kinematics of LBAs include high-velocity components detected in the wings of [\ion{O}{iii}] and Balmer lines, which we interpret as signatures of ionized outflows. 
Examining the relations between the derived outflow properties and the global characteristics of their host galaxies provides valuable clues about feedback processes. 

Figure~\ref{fig:eta_mass_Sigma} (left) shows the relation between $\eta$ and stellar mass for star-forming galaxies at low \citep{Marasco_2023,Mcquin_2019} and high redshift \citep{Concas_2022, llerena2023ionized, weldon2024, Carniani2024, Saldana-Lopez2025, Xu2025} from deep spectroscopic surveys. 
Although the stellar-mass range of the LBA sample is narrow, they are indistinguishable from LBGs at $z \sim 2$–3 and follow the same trend exhibited by low-mass LBGs at $z \gtrsim 3$–9. 

Overall, LBA results agree with those for high-$z$ galaxies ($z \sim 3$) by \citet{llerena2023ionized} based on high-resolution long-slit spectra, including LBGs from \citet{Llerena2022} and strong Ly$\alpha$ emitters \citep{Amorin2017}. 
In the stellar-mass range $M_{\star} \sim 10^{9}$–$10^{10}$\,M$_{\odot}$, our findings are also consistent with \citet{weldon2024}, who used composite MOSDEF spectra of LBGs at $z \sim 1.4$–3.8 (medium-resolution, unresolved). In contrast, our LBAs show larger loading factors at fixed stellar mass than the stacked IFU spectra of main-sequence galaxies at $z \sim 2$ by \citet{Concas_2022}. 
While the $\eta$–$M_{\star}$ trend is similar, the offset likely arises from methodological differences such as lower-resolution IFU data, spectral stacking, or distinct separation of virial and non-virial components in line profiles. 
At higher redshifts, we find good agreement with the trends reported by \citet{Carniani2024} and \citet{Saldana-Lopez2025} using JWST/NIRSpec R2700 spectra, showing a continuous sequence from high- to low-mass, strongly star-forming galaxies with increasing $\eta$. 
Similarly, \citet{Xu2025} and \citet{Cooper2025} derived $\eta$ values between 0.1 and 10 for low-mass galaxies at $z > 3$, displaying comparable trends and large scatter with stellar mass.

Compared with local samples, our results follow the relation found at lower masses ($M_{\star} \sim 10^{7}$–$10^{9}$\,M$_{\odot}$) by \citet{Mcquin_2019}, confirming that low-mass star-forming galaxies strongly influence the outflow mass-loading factor. 
However, our sample shows an offset relative to nearby dwarf galaxies observed with spatially resolved MUSE data by \citet{Marasco_2023}, who report a different trend and generally lower $\eta$. As for the comparison with \citet{Concas_2022}, such differences likely stem from the diverse methodologies used to derive outflow properties \citep[see also][]{Saldana-Lopez2025}.

In Fig.~\ref{fig:eta_mass_Sigma}, we also include predictions from the IllustrisTNG \citep{Nelson_2019} and FIRE-2 \citep{Pandya_2021} simulations, rescaled for comparison with the data. 
IllustrisTNG predicts a characteristic decrease in $\eta$ between $10^{8.5} < M_{\star}/M_{\odot} < 10^{10.5}$, followed by a sharp rise for more massive systems, whereas FIRE-2 shows a monotonic decline from low to high stellar mass, extending down to $M_{\star} \sim 10^{7}$\,M$_{\odot}$. 
The observed LBA and LBG trends are broadly consistent with the predicted higher $\eta$ in low-mass systems due to their shallower potential wells \citep[e.g.][]{Nelson_2019, Pandya_2021}, though the scatter remains substantial (0.5–1 dex) in the lowest-mass bins.

The right panel of Fig.~\ref{fig:eta_mass_Sigma} compares outflow rates $\dot{M}_{\mathrm{out}}$ with SFR surface density $\Sigma_{\mathrm{SFR}}$ for the same samples. 
We find a clear relation between more compact galaxies (higher $\Sigma_{\mathrm{SFR}}$) and larger $\dot{M}_{\mathrm{out}}$, in excellent agreement with high-$z$ LBG studies \citep[e.g.][]{llerena2023ionized, weldon2024, Saldana-Lopez2025}. 
Even the local MUSE sample of \citet{Marasco_2023} follows a similar, though lower, normalization.

Following the conclusions of \citet{llerena2023ionized}, a higher $\dot{M}_{\mathrm{out}}$ in denser starbursts may explain the scatter in $\eta$ at fixed $M_{\star}$. 
Using high-resolution spectra of local star-forming galaxies (including some LBAs), \citet{Xu2025a} found that outflow velocities from optical lines reach 60–70\% of those from UV absorption, yielding outflow rates $\sim$0.2–0.5 dex lower \citep[see also][]{Heckman_2011, Borthakur_2014}. 
These results suggest that the discrepancies arise from the different density regimes probed by emission versus absorption diagnostics.

Future studies combining emission- and absorption-line tracers in this and other local analog samples \citep[e.g. green pea galaxies;][]{Amorin2024} will help clarify how density effects contribute to the observed scatter in these scaling relations and their implications for high-redshift studies.

\section{Conclusions}\label{sec6}

We analyzed the ionized-gas kinematics and outflow properties of 14 compact, UV-luminous galaxies at $z<0.2$—the so-called Lyman-break analogs (LBAs)—using deep, high-resolution (R$\sim$8900) VLT/X-Shooter visible spectra. 
Our analysis involved detailed modeling of H$\beta$, [O\,{\sc iii}]$\lambda\lambda$4959,5007\,\AA, H$\alpha$, [N\,{\sc ii}]$\lambda\lambda$6548,6584\,\AA, and [S\,{\sc ii}]$\lambda\lambda$6716,6731\,\AA{} line profiles, adopting a tailored multi-Gaussian fitting approach with the LiMe software \citep{Lime-paper}. 
Our main results are summarized below:

\begin{enumerate}

\item LBAs exhibit complex emission-line profiles that deviate from a single Gaussian model, requiring two to four components for an accurate representation. 
They include one or more narrow components with intrinsic dispersions $\sigma_{\rm int}\!\sim\!20$–70\,km\,s$^{-1}$ and H$\beta$ luminosities of $10^{40}$–$10^{41}$\,erg\,s$^{-1}$, and at least one broad component with $\sigma_{\rm int}\!\sim\!90$–250\,km\,s$^{-1}$. 
For 9 of the 14 galaxies, the broad components are blueshifted ($\Delta V_r\!\sim\!4$–86\,km\,s$^{-1}$) and contribute $\sim$13–60\% of the total flux in both recombination and collisionally excited lines.

\item From the L(H$\beta$)–$\sigma$ relation for giant extragalactic \hii\ regions and \hii\ galaxies, the narrow components trace virialized gas near young star clusters, while the broad components correspond to high-velocity outflows likely driven by stellar winds and SNe. 
Classical diagnostics indicate stellar photoionization as the main ionizing source, although shocks may contribute to the broad emission in some objects.

\item Electron densities from the [S\,{\sc ii}] doublet span $n_e\!\sim\!100$–1400\,cm$^{-3}$, consistent with other UV-bright starbursts. 
Densities for narrow and broad components are often comparable, although in some galaxies the higher $n_e$ of the broad gas suggests pressurized or shock-heated regions produced by stellar feedback or interactions. 
    
\item Using broad-component measurements and a simple outflow model, we derive mass-loss rates of 0.20–2.72\,M$_{\odot}$\,yr$^{-1}$ (mean 1.4\,M$_{\odot}$\,yr$^{-1}$) and mass-loading factors $\eta$\,=\,0.03–0.81 (mean 0.31), consistent with the mean $\eta$\,=\,0.54 found for LBGs at $z\!\sim\!3$ \citep{llerena2023ionized}. 
Combining with results for lower-mass starbursts at low and high redshift, we find a mild increase of $\eta$ toward lower stellar masses and larger $\dot{M}_{\rm out}$ at higher SFR surface densities. 
Low-mass galaxies undergoing intense starbursts thus develop strong outflows with similar properties across redshift, modulated by the compactness of star formation. 
When derived with comparable methods and assumptions, their outflow properties follow the predicted $\eta$–$M_{\star}$ trends from simulations.

\end{enumerate}

In summary, the multi-Gaussian modeling reveals the strong complexity of the ionized-gas kinematics in LBAs. 
Our findings are consistent with those for high-redshift LBGs and emphasize the role of dense, metal-poor starbursts in shaping outflows and stellar feedback in low-mass galaxies. 
This highlights LBAs as key laboratories for investigating star formation and feedback under ISM conditions akin to those at high redshift. 
Future high-dispersion, spatially resolved (IFU) spectroscopy across multiple wavelengths will be essential to link global gas kinematics, multiphase outflow properties, and star formation in low-mass starbursts.

\section*{Data availability}
Tables~\ref{summary_kinematic_1}  and~\ref{tab:appendix2}-\ref{tab:appendix14} are available in electronic form at the CDS via anonymous ftp to cdsarc.u-strasbg.fr (130.79.128.5) or via http://cdsweb.u-strasbg.fr/cgi-bin/qcat?J/A+A/.

\vspace{2mm}

\begin{acknowledgements}
We are grateful to Verónica Firpo for her valuable guidance on emission-line fitting at the early stages of this work. 
We thank Alberto Saldana-Lopez for providing tables with comparison data. 
AL and RA acknowledge support from ANID Fondecyt 1202007. RA acknowledges support of grant PID2023-147386NB-I00 funded by MICIU/AEI/10.13039/501100011033 and by ERDF/EU, from grant PID2022-136598NBC32 “Estallidos8”, and the Severo Ochoa grant CEX2021-001131-S to the IAA-CSIC. MLl acknowledges support from the INAF Large Grant 2022 “Extragalactic Surveys with JWST” (PI L. Pentericci), the PRIN 2022 MUR project 2022CB3PJ3 - First Light And Galaxy aSsembly (FLAGS) funded by the European Union – Next Generation EU, and the INAF Mini-grant "Galaxies in the epoch of Reionization and their analogs at lower redshift" (PI M. Llerena).

\end{acknowledgements}

\bibliographystyle{aa}
\bibliography{thebibliography}

\begin{appendix}

\section{Electron density}
In Table~\ref{tab:density} we present the electron density ($n_e$) values derived for each galaxy from the integrated profile and each Gaussian component. They are derived as described in Section~\ref{sec:density}.

\begin{table}[H]
\centering
\caption{Electron densities (in cm$^{-3}$) with 16th and 84th percentiles.}
\label{tab:density}
\scriptsize
\begin{minipage}[t]{0.48\linewidth}
\raggedright
\begin{tabular}{ll}
\textbf{SDSS001009} \\
N1 & $261_{104}^{736}$ \\
N2 & $272_{90}^{728}$ \\
N3 & $884_{188}^{3709}$ \\
N4 & $866_{344}^{2027}$ \\
Global & $426_{127}^{1079}$ \\
\\
\textbf{SDSS004054} \\
B & -- \\
N & -- \\
Global & $129_{102}^{156}$ \\
\\
\textbf{SDSS005527} \\
B1 & $1416_{1165}^{1719}$ \\
B2 & $471_{387}^{574}$ \\
N  & $529_{454}^{613}$ \\
Global & $471_{387}^{574}$ \\
\\
\textbf{SDSS015028} \\
B & $212_{113}^{337}$ \\
N & $217_{175}^{261}$ \\
Global & $209_{99}^{393}$ \\
\\
\textbf{SDSS020356} \\
B & $325_{212}^{463}$ \\
N & $50_{37}^{77}$ \\
Global & $127_{73}^{191}$ \\
\\
\textbf{SDSS021348} \\
B & $724_{476}^{1083}$ \\
N & $1197_{903}^{1606}$ \\
Global & $887_{637}^{1232}$ \\
\\
\textbf{SDSS032845} \\
B1 & $144_{64}^{275}$ \\
N1 & $512_{431}^{613}$ \\
N2 & $857_{467}^{1564}$ \\
Global & $360_{270}^{475}$ \\
\end{tabular}
\end{minipage}
\hfill
\begin{minipage}[t]{0.48\linewidth}
\raggedright
\begin{tabular}{ll}
\textbf{SDSS035733} \\
B & $154_{73}^{270}$ \\
N & $142_{70}^{243}$ \\
Global & $132_{73}^{202}$ \\
\\
\textbf{SDSS040208} \\
B & -- \\
N & -- \\
Global & $514_{446}^{586}$ \\
\\
\textbf{SDSS143417} \\
B1 & $120_{55}^{241}$ \\
B2 & $67_{41}^{119}$ \\
N1 & $183_{141}^{229}$ \\
N2 & $120_{58}^{222}$ \\
Global & $101_{55}^{158}$ \\
\\
\textbf{SDSS214500} \\
B1 & $365_{179}^{704}$ \\
N1 & $196_{76}^{487}$ \\
N2 & $134_{73}^{208}$ \\
Global & $219_{99}^{393}$ \\
\\
\textbf{SDSS231812} \\
B & $304_{95}^{865}$ \\
N & $178_{87}^{314}$ \\
Global & $365_{179}^{704}$ \\
\\
\textbf{SDSS232539} \\
B1 & $176_{80}^{331}$ \\
N1 & $504_{134}^{2206}$ \\
N2 & $325_{110}^{857}$ \\
Global & $175_{77}^{336}$ \\
\\
\textbf{SDSS235347} \\
B & -- \\
N & -- \\
Global & $101_{55}^{158}$ \\
\end{tabular}
\end{minipage}
\end{table}

\vspace{1em}

\section{Emission-Line gaussian fitting}\label{appendix:emission-lines}
In Tables~\ref{tab:appendix} to \ref{tab:appendix14} we present the multi-component Gaussian modelling of the emission line profiles for individual galaxies in the sample. Format and units are as in Table~\ref{summary_kinematic_1}. However, here we include properties for the $\hb$, [O{\sc iii}] $\lambda$4959$\AA$ and $\SII$ emission lines.

\noindent
\begin{minipage}[t]{0.48\textwidth}
\scriptsize

\begin{threeparttable}
    \caption{Kinematic properties from multi-component Gaussian fitting. Format and units are as in Table~\ref{summary_kinematic_1}.}
    \label{tab:appendix}
    \begin{tabular}{ccccccc}
        \toprule
        \multicolumn{7}{c}{SDSS001009} \\
        \midrule
        $\lambda_0$ $^{a}$ & Ion & Comp. $^{b}$ & $\sigma_{int}$ $^{c}$ & $\Delta v_r$ $^{d}$ & Flux $^{e}$ & EM $^{f}$ \\
        \midrule
        4861 & H$\beta$ & N1 & 38.3 $\pm$ 2.7 & -6.5 $\pm$ 2.0 & 9.1 $\pm$ 0.5 & 51.8 \\
        & & N2 & 71.5 $\pm$ 21.3 & -95.2 $\pm$ 44.0 & 2.5 $\pm$ 0.6 & 14.5 \\
        & & N3 & 68.5 $\pm$ 10.0 & 275.5 $\pm$ 0.0 & 1.3 $\pm$ 0.5 & 7.2 \\
        & & N4 & 52.5 $\pm$ 2.7 & 87.8 $\pm$ 0.0 & 4.6 $\pm$ 0.5 & 26.6 \\
        \midrule
        4959 & [O{\sc iii}] & N1 & 27.8 $\pm$ 2.3 & -4.7 $\pm$ 1.9 & 6.1 $\pm$ 0.7 & 45.7 \\
        & & N2 & 71.7 $\pm$ 15.5 & -78.6 $\pm$ 0.0 & 2.0 $\pm$ 1.0 & 14.9 \\
        & & N3 & 51.0 $\pm$ 13.7 & 201.0 $\pm$ 0.0 & 1.6 $\pm$ 0.8 & 12.0 \\
        & & N4 & 49.0 $\pm$ 9.6 & 81.3 $\pm$ 0.0 & 3.6 $\pm$ 0.8 & 27.4 \\
        \midrule
        6584 & [N{\sc ii}] & N1 & 39.3 $\pm$ 2.7 & -6.5 $\pm$ 2.0 & 5.4 $\pm$ 0.8 & 43.8 \\
        & & N2 & 72.1 $\pm$ 21.3 & -95.2 $\pm$ 44.0 & 1.9 $\pm$ 0.7 & 15.3 \\
        & & N3 & 69.1 $\pm$ 10.0 & 275.5 $\pm$ 0.0 & 1.4 $\pm$ 0.4 & 11.6 \\
        & & N4 & 53.2 $\pm$ 2.7 & 87.8 $\pm$ 0.0 & 3.6 $\pm$ 0.4 & 29.2 \\
        \midrule
        6716 & [S{\sc ii}] & N1 & 28.9 $\pm$ 3.1 & -12.7 $\pm$ 2.9 & 4.0 $\pm$ 0.6 & 28.0 \\
        & & N2 & 70.4 $\pm$ 10.2 & -141.1 $\pm$ 0.0 & 4.8 $\pm$ 0.7 & 33.6 \\
        & & N3 & 70.0 $\pm$ 23.7 & 236.6 $\pm$ 0.0 & 1.77 $\pm$ 0.6 & 12.3 \\
        & & N4 & 55.9 $\pm$ 8.9 & 88.3 $\pm$ 0.0 & 3.76 $\pm$ 0.7 & 26.1 \\
        \midrule
        6731 & [S{\sc ii}] & N1 & 28.9 $\pm$ 3.1 & -12.7 $\pm$ 2.9 & 2.9 $\pm$ 0.4 & 24.9 \\
        & & N2 & 70.4 $\pm$ 10.2 & -141.1 $\pm$ 0.0 & 3.5 $\pm$ 0.5 & 29.8 \\
        & & N3 & 70.0 $\pm$ 23.7 & 236.6 $\pm$ 0.0 & 1.4 $\pm$ 0.5 & 11.9 \\
        & & N4 & 55.9 $\pm$ 8.4 & 88.3 $\pm$ 0.0 & 4.0 $\pm$ 0.5 & 33.5 \\
        \bottomrule
    \end{tabular}
\end{threeparttable}

\begin{threeparttable}
    \caption{Continue from Table~\ref{tab:appendix}.}
    \label{tab:appendix2}
    \begin{tabular}{ccccccc}
        \toprule
        \multicolumn{7}{c}{SDSS004054} \\
        \midrule
        $\lambda_0$ $^{a}$ & Ion & Comp. $^{b}$ & $\sigma_{int}$ $^{c}$ & $\Delta v_r$ $^{d}$ & Flux$^{e}$ & EM$^{f}$ \\
        \midrule
        4861 & H$\beta$ & B & 121.7 $\pm$ 4.8 & -20.64 $\pm$ 2.3 & 20.6 $\pm$ 2.3 & 13.4 \\
        & & N & 49.2 $\pm$ 0.2 & 22.8 $\pm$ 0.1 & 133.2 $\pm$ 1.6 & 86.6 \\
        \midrule
        4959 & [O{\sc iii}] & B & 120.9 $\pm$ 6.0 & -18.5 $\pm$ 2.8 & 39.8 $\pm$ 1.9 & 12.1 \\
        & & N & 49.9 $\pm$ 0.3 & 25.1 $\pm$ 0.2 & 274.8 $\pm$ 1.5 & 87.9 \\
        \midrule
        6584 & [N{\sc ii}] & B & 122.1 $\pm$ 0.0 & -20.6 $\pm$ 0.0 & 7.8 $\pm$ 1.1 & 35.0 \\
        & & N & 50.1 $\pm$ 0.2 & 22.8 $\pm$ 0.2 & 14.5 $\pm$ 0.7 & 65.0 \\
        \midrule
        6716 & [S{\sc ii}] & B & & & & \\
        & & N & 50.1 $\pm$ 0.2 & 22.8 $\pm$ 0.1 & 22.7 $\pm$ 0.3 & 100 \\
        \midrule
        6731 & [S{\sc ii}] & B & & & & \\
        & & N & 50.1 $\pm$ 0.0 & 22.8 $\pm$ 0.0 & 17.1 $\pm$ 0.2 & 100 \\
        \bottomrule
    \end{tabular}
 
\end{threeparttable}

\begin{threeparttable}
    \caption{Continue from Table~\ref{tab:appendix}.}
    \label{tab:appendix3}
    \begin{tabular}{ccccccc}
        \toprule
        \multicolumn{7}{c}{SDSS005527} \\
        \midrule
        $\lambda_0$ $^{a}$ & Ion & Comp. $^{b}$ & $\sigma_{int}$ $^{c}$ & $\Delta v_r$ $^{d}$ & Flux $^{e}$ & EM $^{f}$ \\
        \midrule
        4861 & H$\beta$ & B1 & 223.9 $\pm$ 10.8 & -27.9 $\pm$ 5.8 & 248.4 $\pm$ 46.2 & 32.5 \\
        & & B2 & 120.1 $\pm$ 5.9 & 5.6 $\pm$ 2.5 & 352.1 $\pm$ 18.5 & 46.0 \\
        & & N & 49.9 $\pm$ 1.6 & 28.0 $\pm$ 0.9 & 164.9 $\pm$ 13.0 & 21.5 \\
        \midrule
        4959 & [O{\sc iii}] & B1 & 274.7 $\pm$ 5.6 & 5.3 $\pm$ 2.1 & 194.6 $\pm$ 5.4 & 19.4 \\
        & & B2 & 127.0 $\pm$ 1.6 & 15.5 $\pm$ 0.5 & 534.2 $\pm$ 5.9 & 53.1 \\
        & & N & 49.0 $\pm$ 0.5 & 33.7 $\pm$ 0.3 & 276.8 $\pm$ 2.8 & 27.5 \\
        \midrule
        6584 & [N{\sc ii}] & B1 & 225.8 $\pm$ 16.0 & -58.0 $\pm$ 15.2 & 255.6 $\pm$ 18.2 & 61.6 \\
        & & B2 & 0.0 $\pm$ 0.0 & 0.0 $\pm$ 0.0 & 0.0 $\pm$ 0.0 & - \\
        & & N & 60.7 $\pm$ 4.7 & 18.2 $\pm$ 3.8 & 159.1 $\pm$ 16.7 & 38.4 \\
        \midrule
        6716 & [S{\sc ii}] & B1 & 224.1 $\pm$ 10.8 & -27.9 $\pm$ 5.0 & 49.6 $\pm$ 2.1 & 26.2 \\
        & & B2 & 120.4 $\pm$ 5.9 & 5.6 $\pm$ 2.2 & 99.4 $\pm$ 2.2 & 52.3 \\
        & & N & 50.6 $\pm$ 1.6 & 28.0 $\pm$ 0.8 & 40.9 $\pm$ 0.9 & 21.5 \\
        \midrule
        6731 & [S{\sc ii}] & B1 & 224.1 $\pm$ 0.0 & -27.9 $\pm$ 0.0 & 60.4 $\pm$ 2.3 & 34.7 \\
        & & B2 & 120.4 $\pm$ 0.0 & 5.6 $\pm$ 0.0 & 75.5 $\pm$ 2.4 & 43.4 \\
        & & N & 50.6 $\pm$ 0.0 & 28.0 $\pm$ 0.0 & 38.2 $\pm$ 0.9 & 21.9 \\
        \bottomrule
    \end{tabular}

    \end{threeparttable}

\end{minipage}
\hfill

\begin{minipage}[t]{0.48\textwidth}
\scriptsize

\begin{threeparttable}
    \caption{Continue from Table~\ref{tab:appendix}.}
    \label{tab:appendix4}
    \begin{tabular}{ccccccc}
        \toprule
        \multicolumn{7}{c}{SDSS015028} \\
        \midrule
        $\lambda_0$ $^{a}$ & Ion & Comp. $^{b}$ & $\sigma_{int}$ $^{c}$ & $\Delta v_r$ $^{d}$ & Flux $^{e}$ & EM $^{f}$ \\
        \midrule
        4861 & H$\beta$ & B & 146.8 $\pm$ 2.5 & -9.4 $\pm$ 0.9 & 58.5 $\pm$ 11.3 & 19.9 \\
        & & N & 71.9 $\pm$ 0.5 & 7.0 $\pm$ 0.3 & 236.1 $\pm$ 8.0 & 80.1 \\
        \midrule
        4959 & [O{\sc iii}] & B & 119.6 $\pm$ 6.2 & -12.4 $\pm$ 2.6 & 108.8 $\pm$ 14.0 & 45.5 \\
        & & N & 50.3 $\pm$ 1.7 & 13.6 $\pm$ 0.9 & 130.5 $\pm$ 9.4 & 54.5 \\
        \midrule
        6584 & [N{\sc ii}] & B & 147.1 $\pm$ 0.0 & -9.4 $\pm$ 1.0 & 164.0 $\pm$ 3.5 & 40.8 \\
        & & N & 72.5 $\pm$ 0.5 & 7.0 $\pm$ 0.3 & 113.0 $\pm$ 0.0 & 59.2 \\
        \midrule
        6716 & [S{\sc ii}] & B & 147.1 $\pm$ 2.5 & 0.1 $\pm$ 0.6 & 49.9 $\pm$ 2.2 & 35.8 \\
        & & N & 74.2 $\pm$ 0.8 & -2.6 $\pm$ 0.5 & 89.5 $\pm$ 1.9 & 64.2 \\
        \midrule
        6731 & [S{\sc ii}] & B & 147.1 $\pm$ 0.0 & 0.1 $\pm$ 0.5 & 39.3 $\pm$ 2.1 & 35.6 \\
        & & N & 74.2 $\pm$ 0.8 & -2.6 $\pm$ 0.5 & 71.3 $\pm$ 1.3 & 64.4 \\
        \bottomrule
    \end{tabular}

    \end{threeparttable}

\begin{threeparttable}
    \caption{Continue from Table~\ref{tab:appendix}.}
    \label{tab:appendix5}
    \begin{tabular}{ccccccc}
        \toprule
        \multicolumn{7}{c}{SDSS020356} \\
        \midrule
        $\lambda_0$ $^{a}$ & Ion & Comp. $^{b}$ & $\sigma_{int}$ $^{c}$ & $\Delta v_r$ $^{d}$ & Flux $^{e}$ & EM $^{f}$ \\
        \midrule
        4861 & H$\beta$ & B & 131.9 $\pm$ 1.2 & -2.7 $\pm$ 0.5 & 100.2 $\pm$ 2.6 & 31.3 \\
        & & N & 48.7 $\pm$ 0.2 & -58.8 $\pm$ 2.9 & 219.1 $\pm$ 1.9 & 68.7 \\
        \midrule
        4959 & [O{\sc iii}] & B & 122.2 $\pm$ 2.1 & -6.0 $\pm$ 2.5 & 113.1 $\pm$ 0.0 & 28.9 \\
        & & N & 47.9 $\pm$ 0.3 & 64.8 $\pm$ 0.4 & 277.9 $\pm$ 1.6 & 71.1 \\
        \midrule
        6584 & [N{\sc ii}] & B & 132.2 $\pm$ 0.0 & -2.7 $\pm$ 0.5 & 33.4 $\pm$ 1.6 & 41.5 \\
        & & N & 49.4 $\pm$ 0.2 & 61.2 $\pm$ 0.1 & 47.1 $\pm$ 0.9 & 58.5 \\
        \midrule
        6716 & [S{\sc ii}] & B & 132.2 $\pm$ 1.2 & -2.7 $\pm$ 0.5 & 35.5 $\pm$ 1.7 & 36.5 \\
        & & N & 49.4 $\pm$ 0.2 & 61.2 $\pm$ 0.1 & 61.6 $\pm$ 1.1 & 63.5 \\
        \midrule
        6731 & [S{\sc ii}] & B & 132.2 $\pm$ 0.0 & -2.7 $\pm$ 0.0 & 26.3 $\pm$ 3.0 & 41.2 \\
        & & N & 49.5 $\pm$ 0.0 & 61.2 $\pm$ 0.0 & 47.7 $\pm$ 1.1 & 58.8 \\
        \bottomrule
    \end{tabular}
    \end{threeparttable}   
    
\begin{threeparttable}
    \caption{Continue from Table~\ref{tab:appendix}.}
    \label{tab:appendix6}
    \begin{tabular}{ccccccc}
        \toprule
        \multicolumn{7}{c}{SDSS021348} \\
        \midrule
        $\lambda_0$ $^{a}$ & Ion & Comp. $^{b}$ & $\sigma_{int}$ $^{c}$ & $\Delta v_r$ $^{d}$ & Flux $^{e}$ & EM $^{f}$ \\
        \midrule
        4861 & H$\beta$ & B & 0.0 $\pm$ 0.0 & 0.0 $\pm$ 0.0 & 0.0 $\pm$ 0.0 &  \\
             &          & N & 38.6 $\pm$ 2.9 & 3.5 $\pm$ 2.9 & 10.5 $\pm$ 0.6 & 100 \\
        \midrule
        4959 & [O{\sc iii}] & B & 0.0 $\pm$ 0.0 & 0.0 $\pm$ 0.0 & 0.0 $\pm$ 0.0 &  \\
             &              & N & 0.0 $\pm$ 0.0 & 0.0 $\pm$ 0.0 & 0.0 $\pm$ 0.0 &  \\
        \midrule
        6584 & [N{\sc ii}] & B & 176.8 $\pm$ 3.2 & -58.7 $\pm$ 2.4 & 69.1 $\pm$ 1.8 & 64.6 \\
             &            & N & 42.6 $\pm$ 0.9 & 3.8 $\pm$ 0.8 & 37.9 $\pm$ 1.0 & 35.4 \\
        \midrule
        6716 & [S{\sc ii}] & B & 118.7 $\pm$ 10.5 & -72.8 $\pm$ 2.6 & 6.7 $\pm$ 1.2 & 62.5 \\
             &            & N & 58.1 $\pm$ 6.6 & 10.1 $\pm$ 4.5 & 7.2 $\pm$ 1.9 & 37.5 \\
        \midrule
        6731 & [S{\sc ii}] & B & 118.7 $\pm$ 0.0 & -72.8 $\pm$ 0.0 & 7.6 $\pm$ 1.1 & 59.2 \\
             &            & N & 58.1 $\pm$ 10.5 & 10.1 $\pm$ 4.5 & 7.0 $\pm$ 1.9 & 40.8 \\
        \bottomrule
    \end{tabular}
    \end{threeparttable}

\begin{threeparttable}
    \caption{Continue from Table~\ref{tab:appendix}.}
    \label{tab:appendix7}
    \begin{tabular}{ccccccc}
        \toprule
        \multicolumn{7}{c}{SDSS032845} \\
        \midrule
        $\lambda_0$ $^{a}$ & Ion & Comp. $^{b}$ & $\sigma_{int}$ $^{c}$ & $\Delta v_r$ $^{d}$ & Flux $^{e}$ & EM $^{f}$ \\
        \midrule
        4861 & H$\beta$ & B1 & 125.8 $\pm$ 1.6 & -29.6 $\pm$ 0.9 & 57.6 $\pm$ 14.1 & 37.0 \\
             &          & N1 & 55.3 $\pm$ 0.37 & -21.3 $\pm$ 0.5 & 80.0 $\pm$ 10.6 & 51.2 \\
             &          & N2 & 18.3 $\pm$ 0.66 & 20.9 $\pm$ 0.5 & 18.5 $\pm$ 4.1 & 11.8 \\
        \midrule
        4959 & [O{\sc iii}] & B1 & 119.4 $\pm$ 5.1 & -28.0 $\pm$ 2.6 & 42.2 $\pm$ 2.1 & 30.5 \\
             &              & N1 & 52.0 $\pm$ 1.8 & -21.3 $\pm$ 1.6 & 82.3 $\pm$ 1.6 & 59.6 \\
             &              & N2 & 21.5 $\pm$ 1.5 & 28.0 $\pm$ 1.0 & 13.7 $\pm$ 0.8 & 9.9 \\
        \midrule
        6584 & [N{\sc ii}]  & B1 & 126.1 $\pm$ 0.0 & -29.6 $\pm$ 1.0 & 35.1 $\pm$ 6.2 & 33.0 \\
             &              & N1 & 55.9 $\pm$ 0.4 & -21.3 $\pm$ 0.4 & 9.5 $\pm$ 2.0 & 57.7 \\
             &              & N2 & 20.2 $\pm$ 0.7 & 20.9 $\pm$ 0.5 & 51.0 $\pm$ 7.9 & 9.4 \\
        \midrule
        6716 & [S{\sc ii}]  & B1 & 123.2 $\pm$ 0.0 & -9.7 $\pm$ 4.9 & 18.8 $\pm$ 0.0 & 31.8 \\
             &              & N1 & 55.9 $\pm$ 0.4 & -21.3 $\pm$ 0.4 & 36.4 $\pm$ 1.0 & 61.7 \\
             &              & N2 & 20.3 $\pm$ 0.7 & 20.9 $\pm$ 0.6 & 3.8 $\pm$ 0.4 & 6.5 \\
        \midrule
        6731 & [S{\sc ii}]  & B1 & 123.2 $\pm$ 0.0 & -9.7 $\pm$ 4.9 & 13.2 $\pm$ 1.2 & 25.8 \\
             &              & N1 & 55.9 $\pm$ 0.0 & -21.3 $\pm$ 0.0 & 33.9 $\pm$ 1.0 & 66.2 \\
             &              & N2 & 20.3 $\pm$ 0.0 & 20.9 $\pm$ 0.0 & 4.0 $\pm$ 0.4 & 7.9 \\
        \bottomrule
    \end{tabular}
    \end{threeparttable}
\end{minipage}
\noindent

\begin{minipage}[t]{0.48\textwidth}
\scriptsize
\begin{threeparttable}
    \caption{Continue from Table~\ref{tab:appendix}.}
    \label{tab:appendix8}
    \begin{tabular}{ccccccc}
        \toprule
        \multicolumn{7}{c}{SDSS035733} \\
        \midrule
        $\lambda_0$ $^{a}$ & Ion & Comp. $^{b}$ & $\sigma_{int}$ $^{c}$ & $\Delta v_r$ $^{d}$ & Flux $^{e}$ & EM $^{f}$ \\
        \midrule
        4861 & H$\beta$ & B & 111.4 $\pm$ 2.4 & 15.1 $\pm$ 0.7 & 45.6 $\pm$ 2.0 & 56.5 \\
             &          & N & 55.4 $\pm$ 1.1 & -21.1 $\pm$ 0.5 & 35.0 $\pm$ 1.5 & 43.5 \\
        \midrule
        4959 & [O{\sc iii}] & B & 119.3 $\pm$ 4.8 & 5.3 $\pm$ 1.4 & 25.7 $\pm$ 2.0 & 53.8 \\
             &              & N & 50.2 $\pm$ 1.6 & -20.3 $\pm$ 0.7 & 22.1 $\pm$ 1.4 & 46.2 \\
        \midrule
        6584 & [N{\sc ii}] & B & 111.7 $\pm$ 0.0 & 15.1 $\pm$ 0.8 & 33.6 $\pm$ 2.7 & 53.9 \\
             &            & N & 56.1 $\pm$ 1.1 & -21.1 $\pm$ 0.6 & 28.7 $\pm$ 1.6 & 46.1 \\
        \midrule
        6716 & [S{\sc ii}] & B & 111.8 $\pm$ 2.4 & 15.1 $\pm$ 0.8 & 25.6 $\pm$ 1.1 & 57.7 \\
             &            & N & 56.1 $\pm$ 1.1 & -21.1 $\pm$ 0.5 & 18.9 $\pm$ 0.7 & 42.3 \\
        \midrule
        6731 & [S{\sc ii}] & B & 111.8 $\pm$ 0.0 & 15.1 $\pm$ 0.0 & 19.5 $\pm$ 0.0 & 57.7 \\
             &            & N & 56.1 $\pm$ 0.0 & -21.1 $\pm$ 0.0 & 14.3 $\pm$ 0.7 & 42.3 \\
        \bottomrule
    \end{tabular}

    \end{threeparttable}

\begin{threeparttable}
    \caption{Continue from Table~\ref{tab:appendix}.}
    \label{tab:appendix9}
    \begin{tabular}{ccccccc}
        \toprule
        \multicolumn{7}{c}{SDSS040208} \\
        \midrule
        $\lambda_0$ $^{a}$ & Ion & Comp. $^{b}$ & $\sigma_{int}$ $^{c}$ & $\Delta v_r$ $^{d}$ & Flux $^{e}$ & EM $^{f}$ \\
        \midrule
        4861 & H$\beta$ & B & 0.0 $\pm$ 0.0 & 0.0 $\pm$ 0.0 & 0.0 $\pm$ 0.0 & \\
             &          & N & 34.0 $\pm$ 0.48 & -17.2 $\pm$ 0.3 & 53.4 $\pm$ 3.3 & 100 \\
        \midrule
        4959 & [O{\sc iii}] & B & 56.5 $\pm$ 4.6 & -20.3 $\pm$ 4.9 & 24.8 $\pm$ 2.7 & 39.0 \\
             &              & N & 33.4 $\pm$ 2.2 & -13.5 $\pm$ 1.4 & 38.8 $\pm$ 1.5 & 61.0 \\
        \midrule
        6584 & [N{\sc ii}] & B & 63.6 $\pm$ 0.0 & -31.0 $\pm$ 2.8 & 8.2 $\pm$ 1.2 & 25.4 \\
             &            & N & 35.1 $\pm$ 0.5 & -17.2 $\pm$ 0.4 & 24.1 $\pm$ 0.9 & 74.6 \\
        \midrule
        6716 & [S{\sc ii}] & B & 0.0 $\pm$ 0.0 & 0.0 $\pm$ 0.0 & 0.0 $\pm$ 0.0 & \\
             &            & N & 26.7 $\pm$ 0.7 & -9.9 $\pm$ 0.7 & 25.2 $\pm$ 0.5 & 100 \\
        \midrule
        6731 & [S{\sc ii}] & B & 0.0 $\pm$ 0.0 & 0.0 $\pm$ 0.0 & 0.0 $\pm$ 0.0 & \\
             &            & N & 34.7 $\pm$ 0.9 & -15.9$\pm$ 0.9 & 23.3 $\pm$ 0.5 & 100 \\
        \bottomrule
    \end{tabular}
    \end{threeparttable}

\begin{threeparttable}
    \caption{Continue from Table~\ref{tab:appendix}.}
    \label{tab:appendix10}
    \begin{tabular}{ccccccc}
        \toprule
        \multicolumn{7}{c}{SDSS143417} \\
        \midrule
        $\lambda_0$ $^{a}$ & Ion & Comp. $^{b}$ & $\sigma_{int}$ $^{c}$ & $\Delta v_r$ $^{d}$ & Flux $^{e}$ & EM $^{f}$ \\
        \midrule
        4861 & H$\beta$ & B1 & 114.9 $\pm$ 7.1 & -85.2 $\pm$ 15.1 & 74.6 $\pm$ 3.7 & 34.7 \\
             &          & B2 & 95.3 $\pm$ 5.9 & 198.0 $\pm$ 10.7 & 14.7 $\pm$ 3.9 & 6.8 \\
             &          & N1 & 44.6 $\pm$ 0.6 & -17.7 $\pm$ 0.3 & 102.8 $\pm$ 3.2 & 47.8 \\
             &          & N2 & 44.6 $\pm$ 2.5 & 93.7 $\pm$ 0.0 & 23.0 $\pm$ 3.0 & 10.7 \\
        \midrule
        4959 & [O{\sc iii}] & B1 & 110.7 $\pm$ 0.0 & -86.0 $\pm$ 20.3 & 14.4 $\pm$ 0.0 & 31.3 \\
             &              & B2 & 80.9 $\pm$ 19.4 & 235.7 $\pm$ 0.0 & 3.7 $\pm$ 0.0 & 8.0 \\
             &              & N1 & 45.4 $\pm$ 2.4 & -14.5 $\pm$ 3.0 & 21.7 $\pm$ 1.1 & 47.2 \\
             &              & N2 & 43.3 $\pm$ 13.5 & 78.3 $\pm$ 0.0 & 6.2 $\pm$ 0.0 & 13.5 \\
        \midrule
        6584 & [N{\sc ii}] & B1 & 115.2 $\pm$ 7.1 & -85.2 $\pm$ 17.8 & 62.7 $\pm$ 5.8 & 19.6 \\
             &             & B2 & 95.7 $\pm$ 0.0 & 198.0 $\pm$ 12.7 & 54.8 $\pm$ 3.7 & 17.2 \\
             &             & N1 & 45.5 $\pm$ 0.6 & -17.7 $\pm$ 0.3 & 165.2 $\pm$ 3.8 & 51.7 \\
             &             & N2 & 45.5 $\pm$ 0.0 & 93.7 $\pm$ 0.0 & 37.0 $\pm$ 4.2 & 11.5 \\
        \midrule
        6716 & [S{\sc ii}] & B1 & 115.2 $\pm$ 7.1 & -85.2 $\pm$ 15.1 & 23.0 $\pm$ 1.2 & 19.8 \\
             &             & B2 & 95.7 $\pm$ 5.9 & 198.0 $\pm$ 10.7 & 28.1 $\pm$ 0.9 & 24.2 \\
             &             & N1 & 45.5 $\pm$ 0.6 & -17.7 $\pm$ 0.3 & 49.2 $\pm$ 0.8 & 42.5 \\
             &             & N2 & 45.5 $\pm$ 2.5 & 93.7 $\pm$ 0.0 & 15.6 $\pm$ 0.7 & 13.5 \\
        \midrule
        6731 & [S{\sc ii}] & B1 & 115.2 $\pm$ 0.0 & -85.2 $\pm$ 0.0 & 15.9 $\pm$ 0.0 & 18.7 \\
             &             & B2 & 95.7 $\pm$ 0.0 & 198.0 $\pm$ 0.0 & 19.1 $\pm$ 0.9 & 22.5 \\
             &             & N1 & 45.5 $\pm$ 0.0 & -17.7 $\pm$ 0.0 & 38.5 $\pm$ 0.8 & 45.3 \\
             &             & N2 & 45.5 $\pm$ 0.0 & 93.7 $\pm$ 0.0 & 11.4 $\pm$ 0.7 & 13.5 \\
        \bottomrule
    \end{tabular}
    \end{threeparttable}
\end{minipage}
\hfill

\begin{minipage}[t]{0.48\textwidth}
\scriptsize
\begin{threeparttable}
    \caption{Continue from Table~\ref{tab:appendix}.}
    \label{tab:appendix11}
    \begin{tabular}{ccccccc}
        \toprule
        \multicolumn{7}{c}{SDSS214500} \\
        \midrule
        $\lambda_0$ $^{a}$ & Ion & Comp. $^{b}$ & $\sigma_{int}$ $^{c}$ & $\Delta v_r$ $^{d}$ & Flux $^{e}$ & EM $^{f}$ \\
        \midrule
        4861 & H$\beta$ & B1 & 117.2 $\pm$ 3.1 & -8.8 $\pm$ 2.1 & 27.1 $\pm$ 2.9 & 17.5 \\
             &          & N1 & 33.9 $\pm$ 1.3 & -61.9 $\pm$ 1.4 & 27.1 $\pm$ 1.0 & 17.5 \\
             &          & N2 & 54.7 $\pm$ 0.9 & 36.7 $\pm$ 0.9 & 100.4 $\pm$ 1.6 & 65.0 \\
        \midrule
        4959 & [O{\sc iii}] & B1 & 103.6 $\pm$ 5.9 & -14.1 $\pm$ 5.8 & 21.4 $\pm$ 3.6 & 30.2 \\
             &              & N1 & 26.5 $\pm$ 1.5 & -67.2 $\pm$ 1.1 & 10.9 $\pm$ 1.1 & 15.4 \\
             &              & N2 & 56.3 $\pm$ 1.8 & 32.9 $\pm$ 1.3 & 38.5 $\pm$ 2.2 & 54.4 \\
        \midrule
        6584 & [N{\sc ii}] & B1 & 117.2 $\pm$ 0.0 & -8.8 $\pm$ 2.5 & 41.6 $\pm$ 4.7 & 29.2 \\
             &             & N1 & 35.1 $\pm$ 1.3 & -61.9 $\pm$ 1.6 & 14.6 $\pm$ 1.7 & 10.5 \\
             &             & N2 & 55.7 $\pm$ 0.0 & 36.7 $\pm$ 1.1 & 86.9 $\pm$ 2.2 & 60.3 \\
        \midrule
        6716 & [S{\sc ii}] & B1 & 117.7 $\pm$ 3.5 & -8.4 $\pm$ 2.5 & 31.3 $\pm$ 3.1 & 35.3 \\
             &             & N1 & 33.0 $\pm$ 3.3 & -57.8 $\pm$ 4.2 & 12.6 $\pm$ 2.2 & 10.7 \\
             &             & N2 & 49.9 $\pm$ 2.1 & 45.0 $\pm$ 2.7 & 42.9 $\pm$ 1.4 & 54.0 \\
        \midrule
        6731 & [S{\sc ii}] & B1 & 117.7 $\pm$ 0.0 & -8.4 $\pm$ 2.5 & 27.1 $\pm$ 2.9 & 39.6 \\
             &             & N1 & 33.0 $\pm$ 3.3 & -57.8 $\pm$ 4.2 & 9.2 $\pm$ 1.1 & 13.5 \\
             &             & N2 & 49.9 $\pm$ 0.0 & 45.0 $\pm$ 0.0 & 32.1 $\pm$ 1.4 & 46.9 \\
        \bottomrule
    \end{tabular}
\end{threeparttable}

\begin{threeparttable}
    \caption{Continue from Table~\ref{tab:appendix}.}
    \label{tab:appendix12}
    \begin{tabular}{ccccccc}
        \toprule
        \multicolumn{7}{c}{SDSS231812} \\
        \midrule
        $\lambda_0$ $^{a}$ & Ion & Comp. $^{b}$ & $\sigma_{int}$ $^{c}$ & $\Delta v_r$ $^{d}$ & Flux $^{e}$ & EM $^{f}$ \\
        \midrule
        4861 & H$\beta$ & B & 152.8 $\pm$ 2.1 & -2.3 $\pm$ 0.8 & 51.9 $\pm$ 1.9 & 26.7 \\
             &          & N & 63.7 $\pm$ 0.3 & 11.6 $\pm$ 0.2 & 142.2 $\pm$ 1.4 & 73.3 \\
        \midrule
        4959 & [O{\sc iii}] & B & 136.5 $\pm$ 7.5 & 22.8 $\pm$ 3.4 & 48.2 $\pm$ 3.7 & 26.2 \\
             &              & N & 67.1 $\pm$ 1.2 & 5.6 $\pm$ 0.5 & 135.8 $\pm$ 2.0 & 73.8 \\
        \midrule

        6584 & [N{\sc ii}] & B & 153.3 $\pm$ 0.0 & -2.3 $\pm$ 0.8 & 29.5 $\pm$ 1.6 & 32.4 \\
             &             & N & 64.3 $\pm$ 0.3 & 11.6 $\pm$ 0.2 & 61.6 $\pm$ 1.0 & 67.6 \\
        \midrule
        6716 & [S{\sc ii}] & B & 145.7 $\pm$ 8.3 & -8.8 $\pm$ 3.9 & 36.2 $\pm$ 5.2 & 40.0 \\
             &             & N & 62.6 $\pm$ 1.7 & 21.6 $\pm$ 0.9 & 54.4 $\pm$ 3.5 & 60.0 \\
        \midrule
        6731 & [S{\sc ii}] & B & 145.7 $\pm$ 0.0 & -8.8 $\pm$ 3.9 & 26.4 $\pm$ 4.3 & 38.5 \\
             &             & N & 62.6 $\pm$ 1.7 & 21.6 $\pm$ 0.9 & 42.1 $\pm$ 1.9 & 61.5 \\
        \bottomrule
    \end{tabular}
    \end{threeparttable}
    
\begin{threeparttable}
    \caption{Continue from Table~\ref{tab:appendix}.}
    \label{tab:appendix13}
    \begin{tabular}{ccccccc}
        \toprule
        \multicolumn{7}{c}{SDSS232539} \\
        \midrule
        $\lambda_0$ $^{a}$ & Ion & Comp. $^{b}$ & $\sigma_{int}$ $^{c}$ & $\Delta v_r$ $^{d}$ & Flux $^{e}$ & EM $^{f}$ \\
        \midrule
        4861 & H$\beta$ & B1 & 148.6 $\pm$ 3.6 & -33.4 $\pm$ 1.6 & 27.3 $\pm$ 2.2& 30.0\\
             &          & N1 & 27.4 $\pm$ 1.5  & -20.5 $\pm$ 0.5 & 16.1 $\pm$ 1.2& 18.3\\
             &          & N2 & 62.4 $\pm$ 2.1  & -18.2 $\pm$ 0.5 & 44.7 $\pm$ 2.6& 50.7\\
        \midrule
        4959 & [O{\sc iii}] & B1 & 165.5 $\pm$ 12.8 & -30.3 $\pm$ 4.6 & 22.9 $\pm$ 1.9 & 18.2 \\
             &              & N1 & 26.8 $\pm$ 1.0   & -17.2 $\pm$ 0.4 & 31.6 $\pm$ 1.0 & 25.1 \\
             &              & N2 & 69.6 $\pm$ 2.5   & -21.1 $\pm$ 0.6 & 71.5 $\pm$ 2.0 & 56.7 \\
        \midrule
        6584 & [N{\sc ii}] & B1 & 148.8 $\pm$ 0.0 & -33.5 $\pm$ 2.0 & 13.7 $\pm$ 1.5 & 46.0 \\
             &             & N1 & 28.7 $\pm$ 1.5  & -20.5 $\pm$ 0.6 & 3.4 $\pm$ 0.7 & 11.4 \\
             &             & N2 & 63.0 $\pm$ 0.0  & -18.2 $\pm$ 0.6 & 12.7 $\pm$ 1.4 & 42.6 \\
        \midrule
        6716 & [S{\sc ii}] & B1 & 148.9 $\pm$ 3.6 & -33.4 $\pm$ 1.6 & 14.3 $\pm$ 0.8  & 57.9\\
             &             & N1 & 28.8 $\pm$ 1.5  & -20.5 $\pm$ 0.5 & 2.5 $\pm$ 0.5  & 10.1\\
             &             & N2 & 63.1 $\pm$ 2.1  & -18.2 $\pm$ 0.5 & 7.9 $\pm$ 1.0  & 32.0\\
        \midrule
        6731 & [S{\sc ii}] & B1 & 148.9 $\pm$ 0.0 & -33.4 $\pm$ 0.0 & 11.0 $\pm$ 0.8 & 58.8 \\
             &             & N1 & 28.8 $\pm$ 0.0  & -20.5 $\pm$ 0.0 & 1.8 $\pm$ 0.5 & 9.6 \\
             &             & N2 & 63.1 $\pm$ 0.0  & -18.2 $\pm$ 0.0 & 5.9 $\pm$ 1.0 & 31.6 \\
        \bottomrule
    \end{tabular}
\end{threeparttable}

\begin{threeparttable}
    \caption{Continue from Table~\ref{tab:appendix}.}
    \label{tab:appendix14}
    \begin{tabular}{ccccccc}
        \toprule
        \multicolumn{7}{c}{SDSS235347} \\
        \midrule
        $\lambda_0$ $^{a}$ & Ion & Comp. $^{b}$ & $\sigma_{int}$ $^{c}$ & $\Delta v_r$ $^{d}$ & Flux $^{e}$ & EM $^{f}$ \\
        \midrule
        4861 & H$\beta$ & B & 95.1 $\pm$ 3.7 & -28.0 $\pm$ 2.6 & 14.9 $\pm$ 3.9 & 14.5 \\
             &          & N & 50.9 $\pm$ 0.6 & 21.4 $\pm$ 0.3 & 87.8 $\pm$ 3.1 & 85.5 \\
        \midrule
        4959 & [O{\sc iii}] & B & 97.5 $\pm$ 8.2 & 6.0 $\pm$ 0.4 & 53.8 $\pm$ 5.5 & 27.5 \\
             &              & N & 49.8 $\pm$ 1.1 & 21.4 $\pm$ 0.4 & 141.5 $\pm$ 2.6 & 72.5 \\
        \midrule
        6584 & [N{\sc ii}] & B & 0.0 $\pm$ 0.0 & 0.0 $\pm$ 0.0 & 0.0 $\pm$ 0.0 & \\
             &             & N & 51.8 $\pm$ 0.6 & 21.4 $\pm$ 0.4 & 14.1 $\pm$ 0.0 & 100 \\
        \midrule
        6716 & [S{\sc ii}] & B & 95.5 $\pm$ 3.7 & -22.4 $\pm$ 2.1 & 9.8 $\pm$ 0.9 & 31.6 \\
             &             & N & 50.5 $\pm$ 0.0 & 15.8 $\pm$ 0.8 & 21.2 $\pm$ 0.7 & 68.4 \\
        \midrule
        6731 & [S{\sc ii}] & B & 95.5 $\pm$ 0.0 & -22.4 $\pm$ 0.8 & 5.7 $\pm$ 0.8 & 26.1 \\
             &             & N & 50.5 $\pm$ 0.0 & 15.8 $\pm$ 0.8 & 16.1 $\pm$ 0.6 & 73.9 \\
        \bottomrule
    \end{tabular}

    \end{threeparttable}
    
\end{minipage}

\section{Additional figures}

\begin{figure}[!ht]
    \centering
    \includegraphics[scale=0.4]{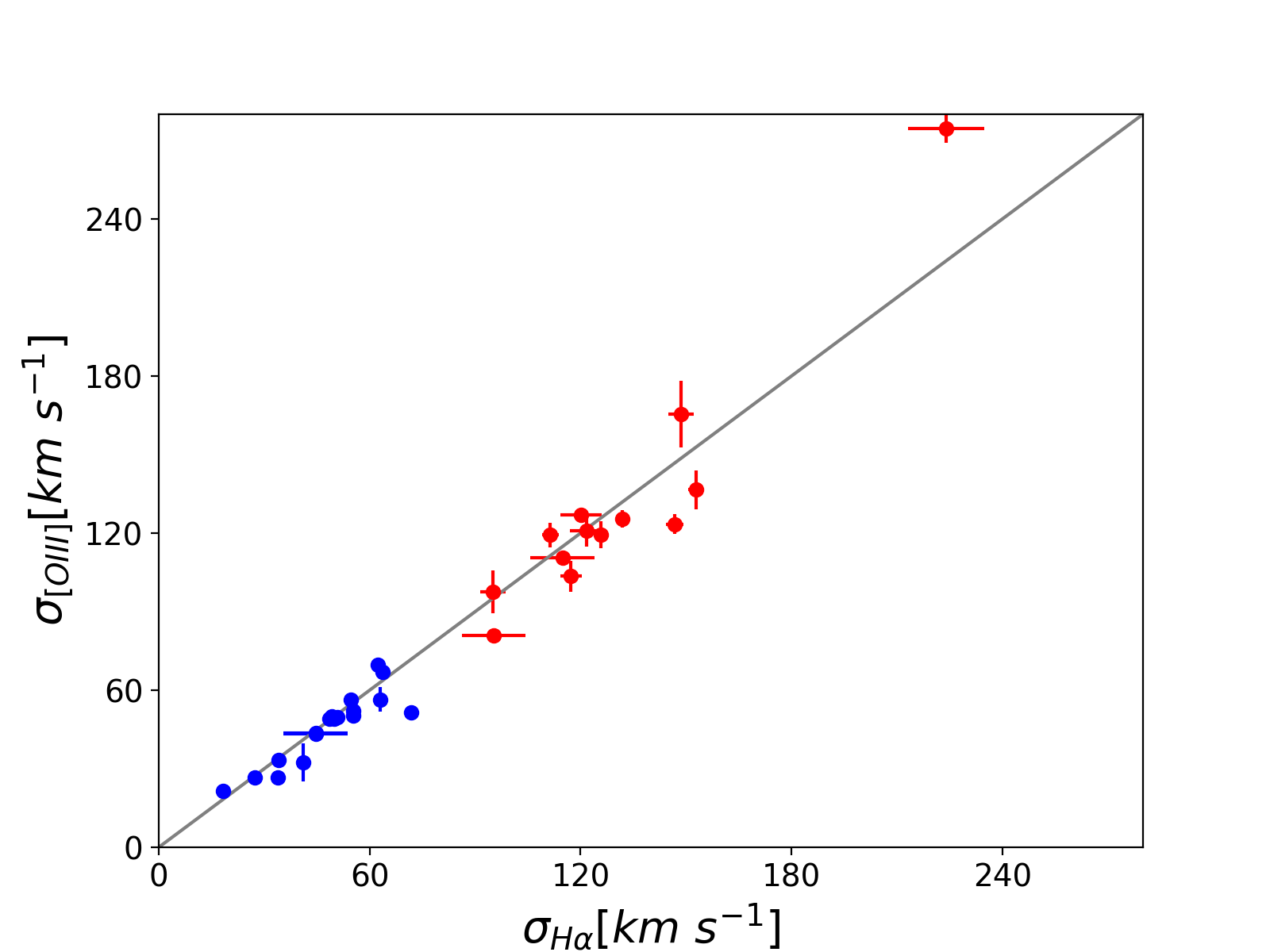}
    \caption{Comparison of $H \alpha$ and $[O{III}]$ valocity dispersion for our multi-component fitting. Blue and red symbols represent narrow an broad Gaussian components. The solid gray line represents the unitary relationship.
    }
    \label{fig:sigma_comparison}
\end{figure}

\begin{figure*}[!ht]
    \centering
    \includegraphics[scale=0.85]{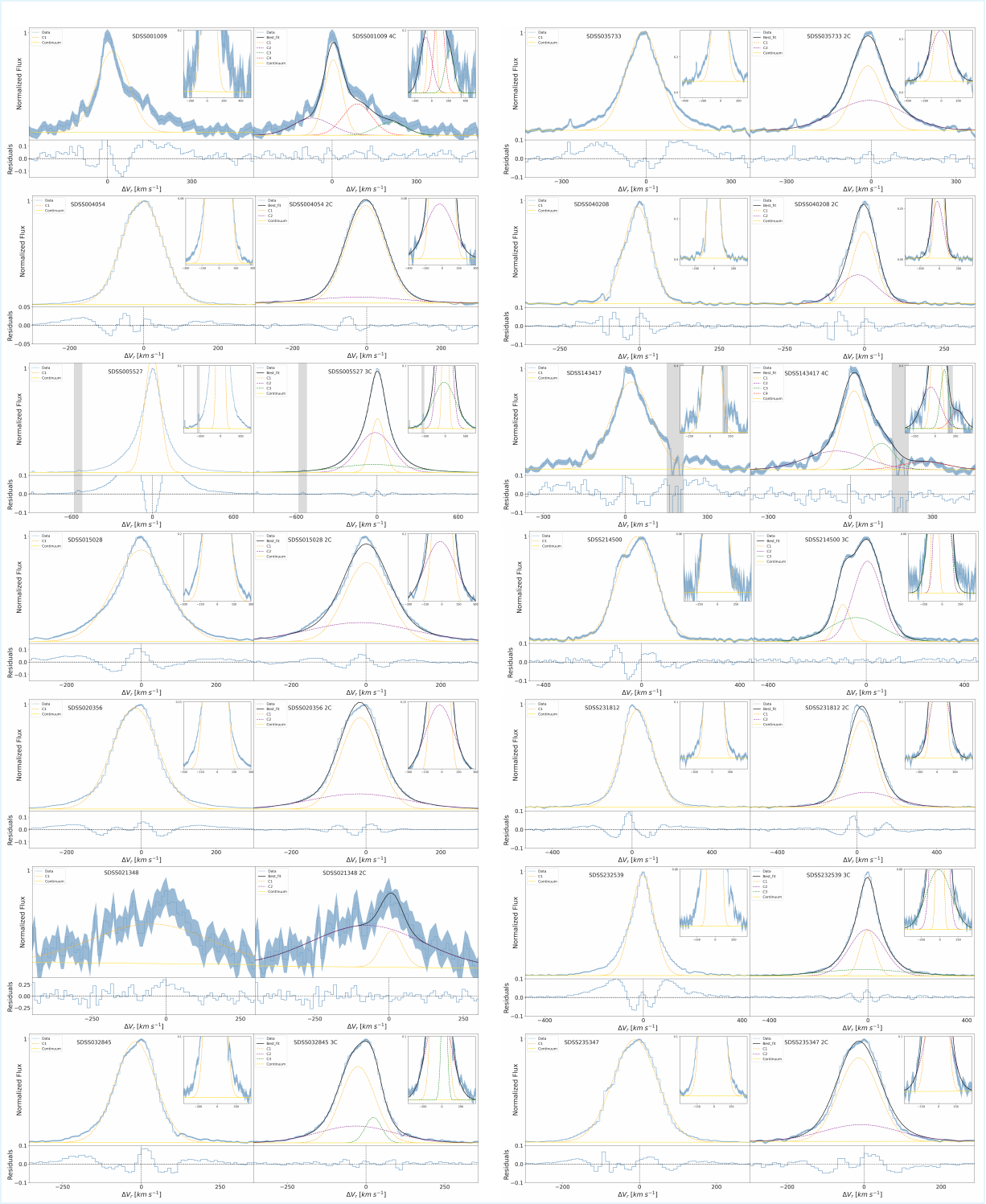}
    \caption{Single Gaussian versus multi-Gaussian line fitting. For each panel, we show the single-component (left) and multi-component (right) model of $[O{III}]$ of each galaxy. Insets, colors, and symbols are as in Fig.~\ref{fig:figura_completa}
    }
    \label{fig:sigma_comparison}
\end{figure*}

\begin{figure*}[!ht]
    \centering
    \begin{minipage}{0.47\textwidth}
        \includegraphics[width=\linewidth]{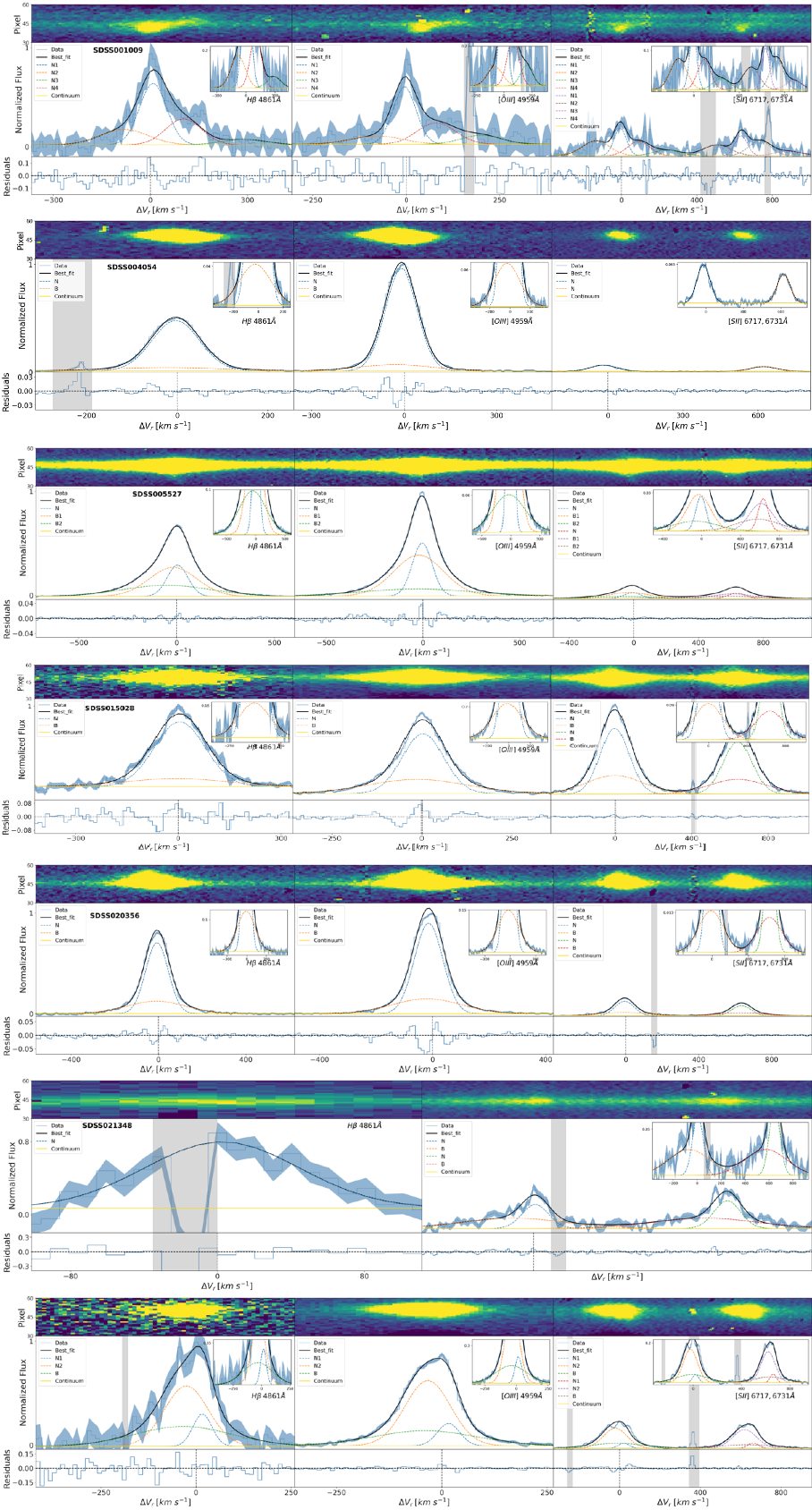}
    \end{minipage}
    \hfill
    \begin{minipage}{0.47\textwidth}
        \includegraphics[width=\linewidth]{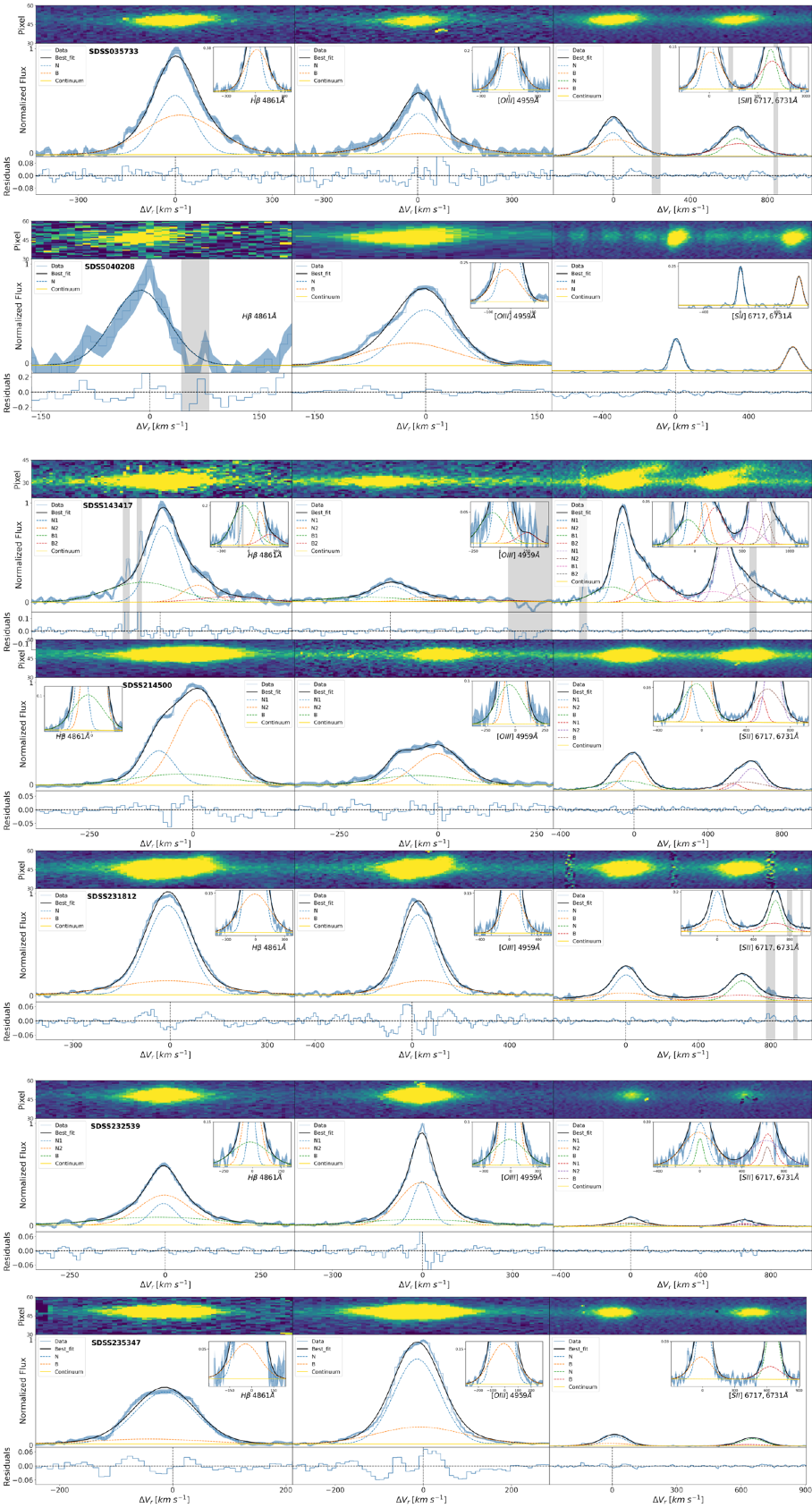}
    \end{minipage}
    \caption{Summary of the multi-Gaussian fitting  of $\hb$, [O{\sc iii}] $\lambda$4959$\AA$, and $\SII$ for the entire sample. The top panels show two-dimensional spectra, with the y-axis in pixel units. Center panels show Gaussian fits for $\hb$ (left), [O{\sc iii}]$\lambda$4959$\AA$ (center) and $\SII$ (right). The bottom panels show the residuals of the fit. The peak emission of each line normalizes the flux. Spectra are shown in light blue (Data). The blue shadow represents the variance spectrum. The black line indicates the fit. The dashed lines show the different fitted components. The yellow line represents the continuum. Grey shadows are flagged regions excluded from fits. The zoom-in insets for the faint line wings are included in the upper-right corner of each plot. }

    \label{fig:figura_completa_faint}
\end{figure*}
\end{appendix}

\end{document}